%% file: main.tex
\newif\iffullversion
\fullversiontrue

\iffullversion
    \documentclass[pra,onecolumn,superscriptaddress]{article}
    \pdfoutput=1
    \usepackage{amsthm}
    \newtheorem{theorem}{Theorem}[section]
    \newtheorem{definition}[theorem]{Definition}
    
    \newtheorem{lemma}[theorem]{Lemma}

    \newtheorem{claim}[theorem]{Claim}
    \newtheorem{construction}[theorem]{Construction}

    \newtheorem{remark}[theorem]{Remark}
    \newtheorem{fact}[theorem]{Fact}
    \newtheorem{assumption}[theorem]{Assumption}

    \usepackage{fullpage}
\else 
	\documentclass[envcountsame,envcountsect]{llncs}

    \newtheorem{construction}[theorem]{Construction}

\fi

\usepackage{amsmath,amsfonts,amssymb}
\usepackage{braket}
\usepackage{graphicx}
\usepackage{algorithm}
\usepackage{tabto}
\usepackage[noadjust]{cite}
\usepackage[dvipsnames]{xcolor}
\usepackage{comment}
\usepackage{footnote}
\makesavenoteenv{algorithm}
\usepackage[backref=page,colorlinks,citecolor=blue,bookmarks=true,final]{hyperref}
\usepackage[bottom]{footmisc}
\usepackage{thm-restate}

\newcommand{\G}{{\mathbb{G}}}
\newcommand{\HH}{{\mathbb{H}}}
\newcommand{\Z}{{\mathbb{Z}}}

\newcommand{\calA}{{\mathcal{A}}}
\newcommand{\calB}{{\mathcal{B}}}
\newcommand{\calD}{{\mathcal{D}}}
\newcommand{\calH}{{\mathcal{H}}}
\newcommand{\calM}{{\mathcal{M}}}
\newcommand{\calR}{{\mathcal{R}}}
\newcommand{\calS}{{\mathcal{S}}}
\newcommand{\calX}{{\mathcal{X}}}

\newcommand{\matI}{{\mathbf{I}}}
\newcommand{\matM}{{\mathbf{M}}}

\newcommand{\vecg}{{\mathbf{g}}}
\newcommand{\vecm}{{\mathbf{m}}}
\newcommand{\vecs}{{\mathbf{s}}}
\newcommand{\vect}{{\mathbf{t}}}
\newcommand{\vecv}{{\mathbf{v}}}
\newcommand{\vecx}{{\mathbf{x}}}
\newcommand{\vecy}{{\mathbf{y}}}

\newcommand{\start}{{\sf Start}}
\newcommand{\act}{{\sf Act}}

\newcommand{\gen}{{\sf Gen}}
\newcommand{\ver}{{\sf Ver}}

\newcommand{\test}{{\sf Test}}

\newcommand{\QFT}{{\sf QFT}}

\newcommand{\negl}{\textnormal{negl}}
\newcommand{\poly}{\textnormal{poly}}

\title{Quantum State Group Actions}

\iffullversion
    \date{}
    \author{
        Saachi Mutreja \\ Columbia University \\ {\tt saachi@berkeley.edu} \and
        Mark Zhandry \\ NTT Research\\{\tt mzhandry@gmail.com}
     }
\else
    \date{}
    \author{}
    \institute{}
\fi

\begin{document}

\maketitle
\begin{abstract}Cryptographic group actions are a leading contender for post-quantum cryptography, and have also been used in the development of quantum cryptographic protocols. In this work, we explore \emph{quantum state} group actions, which consist of a group acting on a set of quantum states. We show the following results:
\begin{itemize}
    \item In certain settings, statistical (even query bounded) security is impossible, analogously to post-quantum classical group actions.
    \item We construct quantum state group actions and prove that many computational problems that have been proposed by cryptographers hold it. Depending on the construction, our proofs are either unconditional, rely on LWE, or rely on the quantum random oracle model. While our analysis does not directly apply to classical group actions, we argue it gives at least a sanity check that there are no obvious flaws in the post-quantum assumptions made by cryptographers.
    \item Our quantum state group action allows for unifying two existing quantum money schemes: those based on group actions, and those based on non-collapsing hashes. We also explain how they can unify classical and quantum key distribution.
\end{itemize}
\end{abstract}
\newpage
\tableofcontents
\newpage
\input{intro}
\input{Prelims}
\input{def}

\input{coset}
\input{hashbased}

\input{generalized_matrix_assumption}
\input{k-wise_ind_expanding_H}

\input{lossyH}

\input{RO}
\input{structuredGMP}

\input{LHS}
\input{qmoney}

\input{qkd}

\bibliographystyle{alpha}
\bibliography{abbrev0,crypto,bib}

\input{appendix}

\end{document}

%% file: intro.tex
\section{Introduction}

Abelian groups have been a fundamental tool in cryptography for almost 50 years, dating back to the pioneering work of Diffie and Hellman~\cite{DifHel76}, and have been used for numerous practical and theoretical results. Unfortunately, Shors quantum algorithm for computing discrete logarithms in abelian groups~\cite{FOCS:Shor94} breaks all of these cryptosystems. Therefore, the looming threat of quantum computers necessitates designing new replacement protocols.

One promising potential replacement is that of abelian \emph{group actions}, originally proposed by~\cite{C:BraYun90}. A group action consists of a group $\G$ acting on a set $\calX$ by a binary operation $*:\G\times\calX\rightarrow\calX$ satisfying $(g+h)*x=g*(h*x)$. At a minimum, a cryptographically-useful group action will satisfy the discrete log assumption: that it is hard to compute $g$ from $x$ and $g*x$. 

Group actions are less-structured versions of groups. Indeed, any abelian group $\mathcal{G}$ can be thought of as a group action, where $\calX=\mathcal{G}$ and $\G=\Z_{|\mathcal{G}|}$, using the action $a*g := g^a$. In this case, the traditional discrete log problem corresponds exactly to the group action discrete log problem. The key difference with group actions is that two set elements cannot be combined in a meaningful way, giving them less structure than plain groups where two elements can be multiplied. Fortunately, this reduced structure prevents applying Shor's algorithm, leading to presumed post-quantum security. Moreover, the conceptual similarity to groups means they can sometimes be used as a drop-in post-quantum replacement, such as for the Diffie-Hellman key agreement protocol. In other cases, a group-based protocol may need adaptation to work with group actions (see for example~\cite{AC:ADMP20} and references therein), or it may be impossible and/or come with inherent costs relative to the classical protocol (e.g.~\cite{EC:BonGuaZha23}).

Cryptographic group actions remain far less studied than their plain group counterparts. For example, isogenies over certain elliptic curves are currently the only family of candidates for post-quantum abelian group actions. Worse, even for ``ideal'' group actions, we do not know if discrete logarithms are hard. The case becomes even less clear for more complicated problems like the linear hidden shift (LHS) assumption proposed in~\cite{AC:ADMP20}. It may be for these assumptions there is a generic quantum attack that simply breaks them on all possible group actions. Contrast to the classical plain group setting, where we know that almost all of the community's favorite problems on groups are at a minimum classically hard for ideal groups~\cite{EC:BonBoyGoh05}, meaning no generic attacks are possible. This leaves the use of group actions for cryptography somewhat precarious.

\subsection{Our Work.} 

In this work, we explore the notion of a \emph{quantum state} group action, which roughly is a group action where the set $\calX$ of classical set elements is replaced with a set of quantum states. Our aim is two-fold:
\begin{itemize}
    \item We position quantum state group actions as a useful concept for exploring \emph{quantum} protocols built from group actions, with potential advantages over classical group actions.
    \item In the setting of quantum attacks on classical cryptosystems, quantum state group actions are not directly relevant as they cannot be implemented in a classical protocol. However, we make the case that understanding the security of quantum state group actions helps improve our understanding of the classical counterpart, which is currently lacking.
\end{itemize}

\noindent We now discuss our contributions in slightly more detail.

\paragraph{Definitions.} The notion of quantum state group action was briefly discussed in~\cite{ITCS:Zhandry24a} in the context of quantum money. We give a formal and precise treatment, identifying certain desirable properties of a quantum state that were implicitly assumed in~\cite{ITCS:Zhandry24a}.

\paragraph{A Hash-based Construction.} We propose a construction of a quantum state group action from hash functions. The set of states are, up to normalization, simply \[|\psi_g\rangle=\sum_x \omega_N^{H(x).g} |x\rangle\] The additive group $\Z_N$ acts on these by \begin{equation}\label{eq:hashbased}h*|\psi_g\rangle=\sum_x \omega_N^{H(x).g} (\omega_N^{H(x).h}|x\rangle)=|\psi_{g+h}\rangle\end{equation}

We prove that a wide class of cryptographic assumptions -- including variants of Decisional Diffie-Hellman (DDH) and the Linear Hidden Shift (LHS) assumption~\cite{AC:ADMP20} -- hold on our group action, for an appropriate choice of hash and for appropriate parameterizations of these assumptions. In some regimes we can even show unconditional hardness when using a $k$-wise independent hash. In other regimes, we prove security in either the quantum random oracle or based on lossy functions (without a trapdoor).

These results show that quantum state group actions can, at least in some regimes, be thought of as a Minicrypt primitive which can be built from simple low-structure objects. Contrast to classical group actions which are considered Cryptomania primitives since they imply public key encryption and require the hardness of highly structured objects from number-theory. Thus, in the setting of quantum cryptographic protocols, quantum group actions may offer a path toward achieving cryptography from milder assumptions. 

We also explain in Section~\ref{sec:motivation} below how the hardness of our quantum state group actions helps improve our confidence in the security of \emph{classical} group actions against quantum attacks, which has remained a challenging open question.

\paragraph{An ``attack'' in the many-copy regime.} Our security proofs only hold when the adversary gets a bounded number of copies of each element. We complement this with a query-efficient but computationally inefficient quantum attack on discrete logarithms -- and hence any reasonable assumption. This attack applies to \emph{any} quantum state group action meeting certain natural requirements. See Section~\ref{sec:coset}. This shows that our results likely cannot extend to the case where many copies of each element are given out.

\paragraph{Unifying quantum money.} Recently, ~\cite{ITCS:Zhandry24a} shows how to construct public key quantum money~\cite{CCC:Aaronson09} from classical abelian group actions, and this construction readily translates to quantum state group actions. We can then attempt to instantiate with our hash-based construction. Unfortunately, there is one feature of group actions that~\cite{ITCS:Zhandry24a} needs which our provably-secure hash-based constructions do not have: the ability to recognize set elements.

It turns out that the resulting construction is \emph{identical} to an earlier quantum money construction of~\cite{EC:Zhandry19b} using something called a non-collapsing hash function. Non-collapsing hashes can be seen as exactly giving the ability to recognize set elements in our hash-based quantum group action. Thus, quantum group actions unifty these two very different looking quantum money schemes.

\paragraph{Unifying key distribution.} The original Diffie-Hellman protocol~\cite{DifHel76} solves the key distribution problem using the computational intractability of certain problems. \emph{Quantum} key distribution solves the same task, but using entirely different principles, namely the rules of quantum information. We show a protocol based on group actions and inspired by Diffie-Hellman which recovers both classical and quantum key distribution, depending on whether the group action is classical and computationally secure, or is quantum and has statistical security. This protocol does not seen to offer any benefits over existing protocols. But we still find it interesting that (quantum) group actions allow for unifying these two very different concepts.

\subsection{Motivation and Additional Discussion}\label{sec:motivation}

\paragraph{Justifying Computational Assumptions.} We now explain how our results, while not directly applicable to the classical group actions used in post-quantum cryptography, nevertheless help justify their use in post-quantum protocols.

Group actions can be thought of as similar to groups, except with less structure. Importantly, this lack of structure means that Shor's algorithm does not work on group actions. Fortunately, there is still enough structure for to implement come cryptographic protocols, such as Diffie-Hellman key agreement. This combination of presumed quantum resistance and cryptographic utility has may group actions one of the leading potential tools for a post-quantum world, with numerous works exploring how to use group actions in the design of cryptosystems resistant to quantum attack. On the other hand, group actions also may be used to achieve novel \emph{quantum} protocols for tasks that are impossible classically, such as proofs of quantumness~\cite{TCC:AlaMalRah22} and quantum money~\cite{ITCS:Zhandry24a}.

Unfortunately, our understanding of the security of group actions is far less developed than that of classical groups. In the classical group and classical attack, a popular formula for justifying the security of cryptoststems is the following: first, prove the security of the cryptosystem under the assumed hardness of some computaitonal problem on the group. There are a vast number of such hard problems, but common examples include discrete logarithm problem or (decisional) Diffie-Hellman. The second step is then to attempt to justify the hardness of this problem. This cannot be accomplished unconditionally, since the current state of complexity theory does not allow for such unconditional lower-bounds. However, we can show for most assumptions of interest that they hold \emph{generically}: that is, that there is no attack that works by only making black box use of the group. This model of adversaries is called the Generic Group Model (GGM)~\cite{Nechaev94,EC:Shoup97,IMA:Maurer05}. Here, the adversary only interacts by making queries to the group. We can then give query complexity lower-bounds for any algorithm solving the hard problem; these query complexity lower bounds then lower bound the overall running time of generic algorithms. Such a generic lower bound is not a full proof of security. But it serves as a sanity check that at least there is not some simple attack which breaks the assumption/scheme.

When moving to group actions and quantum attacks, the first part of this recipe -- proving security from some computational assumption -- is still standard practice. However,  the second part -- justifying the generic hardness of the assumption -- is problematic. It is straightforward to define a Generic Group Action Model (GGAM), as has been done in several recent works~\cite{AC:MonZha22,EC:BonGuaZha23,EPRINT:OrsZan23,PKC:DHKKLR23,ITCS:Zhandry24a}. However, the crucial problem is that even in the GGAM, unconditional lower-bounds are not possible. This is because there exist algorithms with polynomial query complexity (though super-polylnomial run-time) that break any of the standard cryptographic assumptions on group actions~\cite{EttHoy00,EttHoyKni04}. Given that even unconditional lower-bounds are not possible, how can we reason about new hardness assumptions? For example, given a new hardness assumption, perhaps it is actually hard, or perhaps it's actually easy. And in the case that it is easy, perhaps there is even a simple attack that just performs a sequence of simple group action operations. 

A minimal first step would be to prove the generic hardness of the assumption against classical attacks, which is feasible in the GGAM for all of the assumptions on group actions appearing in the literature. However, this is unsatisfying as such hardness proofs also apply for plain groups, which we know are quantumly easy.

We therefore propose a different route, which is to prove unconditional generic hardness for the analogous assumption on \emph{quantum state} group actions. Our hardness results for quantum state group actions in the random oracle model show that, for quantum state group actions, unconditional query complexity lower-bounds are possible. Moreover, our hardness result covers many of the assumptions cryptographers are interested in, such as DDH or LHS. Our hardness result shows that there is no trivial quantum algorithm for these assumptions. We note that our results still do not fully justify these assumptions on classical group actions, since for example Kuperberg's algorithm applies to classical group actions but not quantum state group actions in the single-copy setting. Nevertheless, it is at least a more convincing justification than what was previously known -- which in the quantum setting was very little.

\begin{remark}The discrete logarithm problem on abelian group actions is closely related to the Dihedral Hidden Subgroup Problem (DHSP), and abelian group action discrete logs reduce to DHSP. DHSP is widely studied and is plausibly hard. However, the relationship is only known in one direction, and it is plausible that group action discrete logarithms could be easy even if DHSP is hard. 
\end{remark}

\paragraph{Quantum Cryptography From Mild Assumptions.} Most cryptographic tasks only achieve computational security, meaning security only holds against a bounded adversary. As a consequence, all such schemes rely on un-proven computational assumptions. A common distinction is made between symmetric key primitives (aka Minicrypt) and public key primitives (aka Cryptomania)~\cite{Impagliazzo95}. Symmetric key primitives can be based on mild assumptions such as a one-way function, whereas public key primitives typically require stronger assumptions.

Quantum cryptography offers the possibility of improving the situation, sometimes eliminating the need for computational security at all, such as in quantum key distribution. But for most tasks, even quantum protocols require computaitonal security, yet a recent line of work has shown that the underlying assumptions can often be much milder (e.g.~\cite{EC:GLSV21,C:BCKM21b,ITCS:BraCanQia23,C:KMNY24}).

Our work shows that quantum state group actions may be a useful tool in this push for milder assumptions. Whereas classical group actions give public key encryption and are therefore considered strong tools, in some parameter regimes quantum state group actions require no assumptions at all. In other regimes we base security on random oracles or trapdoor-less lossy functions. These are symmetric key primitives that do not appear to need public key cryptography.

\paragraph{Exponential Security.} Our hash-based scheme in the random oracle model achieves \emph{exponential} security. This is in contrast to classical group actions, which can only achieve at best sub-exponential security due to Kuperberg's algorithm~\cite{Kuperberg05}. This shows in particular that any generic attack on quantum state group actions must take exponential time, and means that standard-model quantum state group actions can potentially achieve exponential security. This stronger security would be advantageous, and may be worth the cost of using quantum communication in certain settings, say if the protocol is quantum anyway.

\subsection{Concurrent And Independent Work}

In a concurrent and independent work,~\cite{MorXag24} also study quantum state group actions, though their notion is somewhat different than ours: instead of being able to act by arbitrary group elements, they only ask for the ability to evaluate the action of randomly selected elements from some distribution over the the unitary group. Thus, the version of quantum state group action considered in~\cite{MorXag24} has a lot less structure than ours. On one hand, this makes their notion potentially easier to instantiate, and they give several candidate instantiations that would not satisfy our stronger requirements. On the other hand, their notion seems to significantly depart from the classical group action abstraction, which allows for acting by arbitrary group elements.

The main result of~\cite{MorXag24} is to show that quantum state group actions give a quantum analog of a pseudorandom function called a pseudorandom function-like state generator; in contrast our main result is to use quantum group actions as a conceptual tool to argue about post-quantum group action assumptions and to unify existing concepts.~\cite{MorXag24} also several candidates from very weak tools (namely, tools that do not seem to imply one-way functions), where as we give a construction from mild-but-not-as-mild tools with provable security. Finally,~\cite{MorXag24} focus on the more general non-abelian case, whereas we consider mainly the abelian case (though our results on coset sampling apply in both cases).

\subsection{Technical Overview}

\paragraph{Defining quantum state group actions (Section~\ref{sec:defs}).} Zhandry~\cite{ITCS:Zhandry24a} briefly define quantum state group actions in the context of quantum money. Here, we identify two important properties of a quantum state group action that were implicitly assumed in~\cite{ITCS:Zhandry24a}:
\begin{itemize}
    \item {\bf Junk-free.} Suppose the procedure for computing $g*|\psi\rangle$ from $g$ and $|\psi\rangle$ additionally produced some ``junk'' state $|\tau_g\rangle$. This would be problematic for any algorithm which computes the group action in super-position over group elements, since the junk states would become entangled with the group elements and de-cohere the resulting state. We therefore expect a group action to be ``junk-free'', and the only output of the procedure is $g*|\psi\rangle$. Junk-free group actions are necessary, for example, to generalize the group-action quantum money scheme of~\cite{ITCS:Zhandry24a} to quantum state group actions.
    \item {\bf Orthogonal.} The next property we consider is orthogonality, which states that the set of quantum states is orthogonal.~\cite{ITCS:Zhandry24a} insists on perfect orthogonality, but we observe that orthogonality may or may not be necessary for applications. We also relax perfect orthogonality to an approximate version, which requires some care. Note that the analysis in~\cite{ITCS:Zhandry24a} requires (approximately) orthogonal group actions, as otherwise some states in the analysis would not be normalized and may even have norm 0.
\end{itemize}
In this work, we focus exclusively on junk-free quantum state group actions, but we consider both orthogonal and non-orthogonal group actions. 

We define a family of matrix assumptions parameterized by a matrix $\matM\in\Z_N^{n\times m}$. The assumption considers two distributions: the first outputs $|\psi_{g_1}\rangle|\psi_{g_2}\rangle\allowbreak\cdots|\psi_{g_n}\rangle$ where $\vecg=(g_1,\cdots,g_n)$ is sampled as $\matM\cdot\vecs$ for a random ``secret vector'' $\vecs\in\Z_N^m$. The second distribution samples $\vecg$ as a uniform random vector. The assumption insists that the two distributions are indistinguishable, given $\matM$ but not $\vecs$ or $\vecg$. We call this the Generalized Matrix Problem (GMP).

This definition encompasses essentially any assumption in the cryptographic group action literature. For example, the decisional Diffie-Hellman (DDH) assumption, when specialized to quantum state group actions, assumption insists on $|\psi_{g_1}\rangle|\psi_{g_2}\rangle|\psi_{g_1+g_2}\rangle$ is indistinguishable from a tuple of random elements. This is a special case of our assumption with $\matM=\left(\begin{array}{cc}1&0\\0&1\\1&1\end{array}\right)$. By generalizing our assumption to non-uniform distributions over secrets $\vecs$, we also obtain the Linear Hidden Shift (LHS) assumption~\cite{AC:ADMP20}, where $\matM$ is a uniformly random matrix and $\vecs$ is a random 0/1 vector. We can also obtain the \emph{extended} LHS assumption of~\cite{TCC:AlaMalRah22} by setting $\matM$ to be a structured matrix. See Definition~\ref{def:GMP} and Section~\ref{sec:defsecurity} for a formal definition and more detail on the examples.

\paragraph{Coset sampling attacks (Section~\ref{sec:coset}).} Classical group actions are subject to quantum coset sampling attacks. For example, given $x_0$ and $x_1=g*x_0$, the attacker prepares a uniform superposition $\sum_{b\in\{0,1\},h\in\G}|b,h\rangle$, and then applies the group action in superposition to compute
\begin{equation}\label{eq:overviewsample}\sum_{b\in\{0,1\},g\in\G}|b,h,h*x_b\rangle\end{equation}
Then the attacker measures the final register, obtaining $y*x_0$ for a random unknown $y$. The remaining superposition then collapses to
\[|\phi_u\rangle=|0,y\rangle+|1,y-g\rangle\]
It is shown in~\cite{EttHoy00,EttHoyKni04} that a polynomial number of samples of $|\psi_y\rangle$ contain enough information about $g$ to (inefficiently) recover $g$.

We show how to generalize this attack to \emph{orthogonal} group actions. Roughly, we hope to be able to use the quantum states $|\psi_0\rangle$ and $|\psi_1\rangle=g*|\psi_0\rangle$ to create an analog of Eq~\ref{eq:overviewsample} where $h*x_b$ is replaced with $h*|\psi_b\rangle$:
\begin{equation}\label{eq:overviewsample2}\sum_{b\in\{0,1\},h\in\G}|b,h\rangle (h*|\psi_b\rangle)\end{equation}
Suppose we can construct such a state. Suppose we measure the third register in the basis containing all of the states $h*|\psi_0\rangle$. The result is that we obtain $y*|\psi_0\rangle$ for an unknown random $y$, and the remaining registers collapse exactly to $|\phi_y\rangle$. 

Now, in general we cannot efficiently measure in the basis containing the states $h*|\psi_0\rangle$. However, we actually do not need to, since we do not actually need the measurement outcome, just the state collapse. Simply discarding the final register has the same effect, giving a sample of $|\psi_u\rangle$. We can repeat several times (using fresh copies of $|\psi_0\rangle,|\psi_1\rangle$ each time) to get several samples.

We return to constructing the state in Eq~\ref{eq:overviewsample2}, and see that there is a problem. If we have a few copies of $|\psi_0\rangle$ and $|\psi_1\rangle$, we have to use up a copy to construct the state in Eq~\ref{eq:overviewsample2}, and which state we use is determined by $b$. If done naively, our remaining local copies will therefore determine $b$ -- simply look at which of $|\psi_0\rangle,|\psi_1\rangle$ has fewer copies left -- meaning the local copies are entangled with our desired state. This breaks the attack. We fix the attack by preparing the state
\begin{equation}\label{eq:overviewsample3}\sum_{b\in\{0,1\},h\in\G}|b,h_0,h_1,\rangle (h_0*|\psi_b\rangle) (h_1*|\psi_b\rangle)\end{equation}
This uses up exactly one copy of each of $|\psi_0\rangle,|\psi_1\rangle$. Now measure the final two registers (or really, just discard them). The result is $y_0*|\psi_0\rangle$ and $y_1*|\psi_1\rangle$. A simple analysis then shows that the remaining registers collapse to the state $|0,y_0,y_1\rangle+|1,y_0-g,y_1+g\rangle$. This has exactly the form of $|\phi_{(y_0,y_1)}\rangle$ for an unknown discrete log $(g,-g)\in\G^2$. By applying~\cite{EttHoy00,EttHoyKni04} to this larger group, we obtain $(g,-g)$ and hence $g$.

\paragraph{Our Hash-Based Construction (Section~\ref{sec:hashbased}).} Next, we turn to our hash-based construction described in Equation~\ref{eq:hashbased}. It is easy to see that the construction is junk-free. We also give a technical lemma, showing that if $H$ is pairwise independent and sufficiently compressing, then it is orthogonal.

\paragraph{Proving the security of our hash-based construction (Sections~\ref{sec:gmp1} and~\ref{sec:gmp2}).} We now turn to proving the security of our hash-based construction, with the aim of justifying the hardness of the GMP problem. We prove, under certain assumptions about $\matM$ and the distribution over $\vecs$, that the above assumption is indeed true for our hash-based quantum state group action. In this overview, we will focus on the case of DDH.

Recall that the elements in this group action have the form
\[|\psi_g\rangle=\sum_x \omega_N^{H(x).g} |x\rangle\]
Now look at the state $|\psi_{g_1}\rangle|\psi_{g_2}\rangle|\psi_{g_1+g_2}\rangle$, which for our hash-based construction gives
\[\sum_{x_1,x_2,x_3}\omega_N^{H(x_1).g_1+H(x_2).g_2+H(x_3)\cdot (g_1+g_2)}|x_1,x_2,x_3\rangle=\sum_{x_1,x_2,x_3}\omega_N^{f(x_1,x_2,x_3)\cdot(g_1,g_2)}|x_1,x_2,x_3\rangle\]
where we have defined $f(x_1,x_2,x_3)=(H(x_1)+H(x_3),H(x_2)+H(x_3))$. Above, $g_1,g_2$ are random elements in $\Z_N$. We can therefore look at the density matrix where we average over $g_1,g_2$. The result is the mixed state
\[\rho=\sum_{\substack{x_1,x_2,x_3\\x_1',x_2',x_3'}}\sum_{g_1,g_2}\omega_N^{\big(f(x_1,x_2,x_3)-f(x_1',x_2',x_3')\big)\cdot (g_1,g_2)}|x_1,x_2,x_3\rangle\langle x_1',x_2',x_3'|\]
We can now perform the sum over $g_1,g_2$. This has the effect of zeroing out everywhere that $f(x_1,x_2,x_3)-f(x_1',x_2',x_3')\neq 0$, leaving only terms with $f(x_1,x_2,x_3)=f(x_1',x_2',x_3')$. 

It turns out that this is \emph{equivalent} to the density matrix obtained by starting with the uniform superposition over $|x_1,x_2,x_3\rangle$, and measuring $f(x_1,x_2,x_3)$. Suppose then that the function $f$ is injective. Then measuring the output of $f$ is equivalent to measuring the input to $f$, or in particular measuring $x_1,x_2,x_3$. By performing a similar analysis to the state $|\psi_{g_1}\rangle|\psi_{g_2}\rangle|\psi_{g_3}\rangle$ for uniform independent $g_1,g_2,g_3$, we obtain that as long as $f'(x_1,x_2,x_3)=(H(x_1),H(x_2),H(x_3))$ is injective, then the state is equivalent to measuring $x_1,x_2,x_3$. Thus, if we can prove that $H$ satisfies this strong pair of injectivity requirements, we will have shown that the two distributions are indistinguishable.

First, injective $f'$ is equivalent to injective $H$, and a pair-wise independent sufficiently-expanding function will be injective. We show that $f$ is also ``almost'' injective, supposing $H$ is sufficiently expanding and $k$-wise independent for sufficiently large $k$. This injectivity does not hold for all points, since $f(x_1,x_1,x_2)$ collides with $f(x_2,x_2,x_1)$. But we show that these edge cases only incur a small statistical error. Thus, we obtain an $H$ for which DDH \emph{unconditionally} holds.

When $H$ is injective, however, the group action is not orthogonal, since an expanding hash function means the dimension of the elements $|\psi_g\rangle$ is smaller than the size of $\G$. We are also interested in constructing orthogonal quantum-state group actions, and give two such solutions. One uses lossy functions. Basically, in the injective mode, we can make $H$ pairwise independent so the group action will in fact be orthogonal via our technical lemma. For security, we switch to the lossy mode where $H$ behaves as if it is expanding, allowing the above analysis to go through. We also show that $H$ can be set to be a random oracle.

We extend the above security proof to work for any $\matM$ for which the rows are distinct; see Section~\ref{sec:gmp1}. Our analysis puts an upper bound on how tall $\matM$ can be in terms of the size $N$ of the group, which is inherent for statistical security in the expanding $H$ case. We also generalize to some cases of non-uniform $\vecs$. Here, summing over all of the (now weighted) $g_i$ does not exactly zero-out entries of the density matrix $\rho$. We give conditions under which we can prove security. Specifically, if $\matM$ has linearly independent columns and $\vecs$ has sufficiently high min-entropy, security holds.

We show that as long as $\matM$ has linearly independent rows, and the min-entropy of $\vecs$ is high enough, that security holds. In particular, we require the min-entropy to grow with the number of rows. This encompasses a variant of the (extended) LHS assumption where $\vecs$ has high min-entropy instead of being a random 0/1 vector. We also show that, in the specific case that $\matM$ is random and short and wide, that we can take $\vecs$ to be random 0/1 vector. This encompasses the LHS assumption in the small-sample regime; see Section~\ref{sec:gmp2}.

\paragraph{Quantum Money (Section ~\ref{qmoney}).} Zhandry~\cite{ITCS:Zhandry24a} gives a quantum money scheme from classical abelian group actions. The banknote for serial number $h$ is $|\$_h\rangle:=\sum_g \omega_N^{gh}|g*x\rangle$. In~\cite{ITCS:Zhandry24a}, it is shown how to efficiently mint $|\$_h\rangle$ for a random $h$ (but importantly, \emph{not} a chosen $h$), as well as how to learn $h$ from $|\$_h\rangle$.

Let us now see what happens when we plug in our hash-based quantum-state group action. Then
\[|\$_h\rangle=\sum_g\omega_N^{gh}|\psi_g\rangle=\sum_{g,x}\omega_N^{gh+g H(x)}|x\rangle\]
We can then carry out the sum over $g$, which zero's out every term with $h+H(x)\neq 0$. The result is exactly the state
\[|\$_h\rangle=\sum_{x:H(x)=-h}|x\rangle\]
While minting $|\$_h\rangle$ on general group actions requires some work, minting $|\$_h\rangle$ on our hash-based group action is very easy: simply create the uniform superposition over $x$, an measure $H(x)$ to get $-h$; the state then collapses to $|\$_h\rangle$.

This kind of money state actually has already appeared previously in~\cite{EC:Zhandry19b}. Interestingly, for typical hash functions, we expect that cloning $|\$_h\rangle$ is easy: one can just measure the state to get a classical $x$ that is a pre-image of $-h$, and copy $x$ at will. The only way to block this attack is to have a method of distinguishing a classical $x$ from the superposition over pre-images $|\$_h\rangle$. Such a hash function is called non-collapsing~\cite{EC:Unruh16}, and~\cite{EC:Zhandry19b} shows that, indeed, non-collapsing hashes are sufficient for secure quantum money. Unfortunately, little is known about how to construct such hashes. Likewise, there is an ingredient in~\cite{ITCS:Zhandry24a} that we omitted above: namely the ability to recognize set elements. Under our correspondence using our hash-based quantum group action, the ability to recognize set elements is exactly the ability to distinguish classical $x$ from superpositions. Due to lack of space, we defer the details to Appendix~\ref{qmoney}.

\paragraph{Key Distribution (Section ~\ref{qkd}).} As an additional application of quantum group actions, we also present a key distribution protocol from abelian group actions. When instantiated with an information theoretically secure quantum group action, our protocol is an information theoretically secure protocol for quantum key distribution. 

%% file: Prelims.tex
\section{Preliminaries}
\begin{definition}[Minimum Entropy]
The min entropy of a random variable $X$ is
\[
H_{\infty}(X)= \min_{u \in U}{-\log(\Pr[X=u])}= -\log(\max_{u\in U}\Pr[X=u]).
\]
\end{definition}
\begin{definition}[Strong Extractor]
   Let the seed $U_d$ be uniformly distributed on $\{0,1\}^d$. We say that a function $Ext:\{0,1\}^n\times \{0,1\}^d \rightarrow \{0,1\}^m$ is a $(k,\epsilon)$ strong extractor if, for all random variables $X$ on $\{0,1\}^n$ independent of $U_d$ with $H_{\infty}(X)\geq k,$
   \[
   d_{TV}((Ext(X,U_d), U_d),(U_m,U_d))\leq \epsilon
   \]
   where $U_m$ is uniformly distributed over $\{0,1\}^m$ independent of $X$ and $U_d$.
\end{definition}

\begin{definition}[Leftover Hash Lemma]
 Let $X$ be a random variable with universe $U$ and $H_{\infty}(X)\geq k$. Fix $\epsilon>0$. Let $\mathcal{H}$ be a universal hash family of size $2^d$ with output length $m=k-2\log(1/\epsilon)$. Define $
 Ext(x,h)=h(x).$ Then $Ext$ is a strong $(k,\epsilon/2)$ extractor with seed length $d$ and output length $m$.
\end{definition}

%% file: def.tex
\section{Defining Quantum State Group Actions}\label{sec:defs}

\paragraph{Classical Group Actions.} A group action consists of a (multiplicative) group $\G$ that acts on a set $\calX$ via an action $*$. The requirements of a group action are:
\begin{itemize}
    \item {\bf Identity}: if $e$ is the identity in $\G$, then $e*x=x$ for all $x\in\calX$.
    \item {\sf Compatibility}: For any $g,h\in\G$ and $x\in\calX$, $g*(h*x)=(gh)*x$.
\end{itemize}

\paragraph{Quantum Group Action.} Here we define quantum group actions. A quantum group action will consist of:
\begin{itemize}
    \item A family of classical group actions $(\; (\G_\lambda,\calX_\lambda,*_\lambda)\;)_\lambda$ for $\lambda\in\Z^+$, where $|\G_\lambda|,|\calX_\lambda|\leq 2^{O(\lambda)}$. We will usually drop the subscript $\lambda$ from $*$, and just write $g*x$ for $g\in\G_\lambda,x\in\calX_\lambda$. When $\lambda$ is clear from context, we will also abuse notation and often omit the subscript $\lambda$ on $\G_\lambda,\calX_\lambda$, and $*_\lambda$.
    \item A family $\Psi=(\Psi_\lambda)_\lambda$ for $\lambda\in\Z^+$, where $\Psi_\lambda=(|\psi_x\rangle)_{x\in\calX_\lambda}$ is a collection of states indexed by $\calX_\lambda$. The states $|\psi_x\rangle$ lie in a Hilbert space $\calH_\lambda$.
    \item A distinguished starting element $x_\lambda\in\calX_\lambda$
    \item A polynomial-time quantum procedure $\start(1^\lambda)$ which outputs a (potentially mixed) state $\rho$.
    \item A polynomial-time quantum procedure $\act(g\in\G_\lambda,|\psi\rangle\in\calH_\lambda)$, which outputs a (potentially mixed) state $\rho$ over $\calH_\lambda$.
\end{itemize}

\noindent We now discuss the desired properties of a quantum group action.

\begin{definition}[Correctness] A quantum group action as described above is \emph{correct} if there exists a negligible function $\negl(\lambda)$ such that:
\begin{align*}
    T(\;|\psi_{x_\lambda}\rangle\langle \psi_{x_\lambda}|\; , \; \start(1^\lambda)\;)&\leq \negl(\lambda)\\
    T(\;|\psi_{g*x}\rangle\langle\psi_{g*x}|\; , \; \act(g,|\psi_x\rangle\langle\psi_x |)\;)&\leq\negl(\lambda),\;\forall x\in\calX_\lambda,g\in\G_\lambda
\end{align*}
If $\negl(\lambda)=0$, then we say the group action is \emph{perfectly} correct.
\end{definition}
In other words, $\start$ (approximately) produces $|\psi_{x_\lambda}\rangle$, and $\act$ (approximately) implements the group action $*$ over the states $|\psi_x\rangle$. 

\medskip

The next definition considers how $\act$ behaves in terms of side information, or junk, produced. Note that, because the inputs and outputs of $\act$ are quantum, there is no generic mechanism for uncomputing workspace qubits or for leaving a copy of the input intact while computing the output. We here define the notion of ``junk-free'' quantum group actions, which roughly states that the input is consumed in producing the output, and no workspace qubits are produced.
\begin{definition}[Junk-Free] A quantum group action as described above is \emph{junk-free} if $\act(g,|\psi\rangle)$ works as follows:
\begin{itemize}
    \item Based on $g\in\G_\lambda$, $\act$ computes a description of a polynomial-sized quantum circuit $U_g$. $U_g$ acts on $\calH_\lambda\times \calH_{{\sf work},\lambda}$, where $\calH_{{\sf work},\lambda}$ are workspace qubits.
    \item $\act$ then applies $U_g|\psi\rangle|0\rangle$.
    \item $\act$ then measures $\calH_{{\sf work},\lambda}$; if the result is not equal to $0$, then $\act$ outputs a special symbol $\bot$ and aborts.
    \item If $\act$ does not abort, then it outputs the state in $\calH_\lambda$.
\end{itemize}
\end{definition}

\noindent We next give the definitions of \emph{orthogonality} and \emph{regularity}:
\begin{definition}[Orthogonal] A quantum group action is \emph{orthogonal} if, for all $x,y\in\calX_\lambda$ such that $x\neq y$, $\langle\psi_x|\psi_y\rangle=0$.
\end{definition}

\begin{definition}[Regular] A group action $\G,\calX,*$ is \emph{regular} if the map $(g,x)\mapsto (g,g*x)$ is a bijection. A quantum group action as described above is regular if each of the underlying group actions in the group action family is regular.
\end{definition}

\paragraph{Notations and conventions.} We always assume a correct, junk-free, and regular quantum group action. Note that a quantum group action will be specified by $\G=(\G_\lambda)_\lambda$, $\start$ and $\act$, with the other terms being implicitly derived from them. For any given $(\start,\act)$, there may be several possible realizations of the underlying classical group actions and sets $\Psi$. When we say that a group actions satisfies a particular property, we mean that there is \emph{some} realization under which the action satisfies that property. 

We write $g*|\psi_x\rangle$ to denote $\act(g,|\psi_x\rangle)$. By correctness, this approximately gives $|\psi_{g*x}\rangle$. We will abuse notation and write $|\psi_{g*x}\rangle=g*|\psi_x\rangle$. We will typically treat $|\psi_{x_\lambda}\rangle$ and the output of $\start(1^\lambda)$ as equal, as well as $|\psi_{g*x}\rangle$ and $g*|\psi_x\rangle$.

For a regular group action, the set $\calX_\lambda$ is in bijection with the group $\G_\lambda$. Wetherefore take $\calX_\lambda$ to be exactly $\G_\lambda$, and $x_\lambda$ to be $e_\lambda$, the identity in $\G_\lambda$.

A \emph{classical} group action is a group action where the states $|\psi_x\rangle$ are the elements $|x\rangle$ in the computational basis. Such a classical group action is automatically orthogonal. It can also be made junk free: since the inputs and outputs to $\act$ are classical, any workspace can be un-computed via standard techniques.

\subsection{Cryptographic Quantum Group Actions} \label{sec:defsecurity}

\paragraph{Discrete Logarithms.} For cryptographic purposes, there needs to be a hard problem associated with the group action, which is always \emph{at least} the hardness of computing discrete logarithms: computing $g$ given $|\psi_{g*x_\lambda}\rangle$. In the quantum setting, the element $|\psi_{g*x_\lambda}\rangle$ is not necessarily clonable, so we parameterize the problem based on how many copies of $|\psi_{g*x_\lambda}\rangle$ are given.

\begin{definition}[Discrete Logarithm Problem] For a function $\ell=\ell(\lambda)$, the $\ell$-\emph{discrete logarithm} (DLog) problem is hard in a quantum group action if, for every quantum polynomial-time adversary $\calA$, there exists a negligible function $\negl$ such that:
\[\Pr\left[g*x_\lambda=g'*x_\lambda:\begin{array}{rl} g&\gets\G_\lambda\\g'&\gets\calA\left(|\psi_{g*x_\lambda}\rangle^{\otimes \ell(\lambda)}\right)\end{array}\right]\leq\negl(\lambda).\]
The DLog problem is \emph{hard} (without parameterization by $\ell$) if DLog is $\ell$-hard for all polynomials $\ell$. Note that for regular (quantum) group actions, the condition on the left simply becomes $g=g'$.\end{definition}

\paragraph{Generalized Matrix Assumption.} Other assumptions have also been made on classical group actions. A basic example is that of Decisional Diffie Hellman (DDH), which asks that $g*x,h*x,(gh)*x$ is indistinguishable from three random set elements. The DDH assumption is the group-action analog of DDH on plain groups, and is sufficient for key agreement. 

A more complex assumption is the Linear Hidden Shift (LHS) assumption~\cite{AC:ADMP20}. The LHS assumption, very roughly, gives out many samples of the form $(g_1,\cdots,g_k,(\prod_i g_i^{s_i})*x)$, where the $g_i$ are fresh random elements in each sample, while the $s_i$ are random bits shared between all the samples.

Here, we define a generalization of DDH, which we call the Generalized Matrix Problem (GMP). This assumption encompasses DDH, LHS, and potentially many more assumptions on group actions. We also describe the assumption on quantum state group actions. 

We first given some notation. For a matrix $\matM=(M_{i,j})_{i,j}\in\G^{m\times n}$ and vector $\vecv=(v_j)_j\in\Z^n$, define $\matM\cdot\vecv$ to be the vector in $\G^m$ whose $i$th entry is $\prod_j M_{i,j}^{v_j}$. Likewise, if $\matM\in\Z^{m\times n}$ and vector $\vecv\in\G^n$, then $\matM\cdot\vecv$ is the vector in $\G^m$ whose $i$th entry is $\prod_j v_j^{M_{i,j}}$. For a vector $\vecv=(v_i)_i\in\G^m$, let $\vecv*|\psi_x\rangle^m=(v_i*|\psi_x\rangle)_i$.

For a matrix $\matM=(M_{i,j})_{i,j}$, we let $M_i$ denote the $i$th row of $\matM$, namely $M_i=(M_{i,j})_j$. For two matrices $\matM^{(0)},\matM^{(1)}$ of the same height, we say that they have the same \emph{row equality pattern} if, for each $i,i'$, $M^{(0)}_{i}=M^{(0)}_{i'}$ if and only if $M^{(1)}_{i}=M^{(1)}_{i'}$. Note that we can think of vectors as column matrices with a single column. In this way, it makes sense to talk about a matrix and vector having the same row equality pattern.

\begin{definition}[Generalized Matrix Problem]\label{def:GMP} Let $m,n$ be polynomials in $\lambda$. Let $\calM=(\calM_\lambda)_\lambda$ be a distribution over $\G^{m\times n}$ (resp. over $\Z^{m\times n}$) and let $\calS=(\calS_\lambda)_\lambda$ be a distribution over $\Z^{n}$ (resp. $\G^{n}$). Then the $(\calM,\calS)$-Generalized Matrix (GM) Problem is hard in a quantum group action if, for every quantum polynomial-time adversary $\calA$, there exists a negligible function $\negl$ such that:
\[\Pr\left[b'=b:\begin{array}{rl}
    b&\gets\{0,1\}\\
    \matM&\gets\calM_\lambda\\
    \vecs&\gets \calS_\lambda\\
    \vecv^{(0)}&\gets \matM\cdot\vecs\\
    \vecv^{(1)}&\gets {\sf EqPat}_\matM(\G^m)\\
    b'&\gets \calA(\matM,\vecv^{(b)}*|\psi_{x_\lambda}\rangle^m)
\end{array}\right]\leq 1/2+\negl(\lambda).\]
Above, ${\sf EqPat}_\matM(\G^m)$ means the subset of $\G^m$ having the same row equality pattern as $\matM$.\end{definition}

\begin{remark}\label{rem:mixmatch} We can also generalize Definition~\ref{def:GMP} to $\matM$ and $\vecs$ each contain both elements of $\G$ and elements of $\Z$, as long as the entries are positioned so that in the matrix-vector product, we only ever multiply a group element by an element in $\Z$.
\end{remark}

\begin{remark}We can also extend GM problem to the case where $\ell$ copies of each group element are given out. However, this is equivalent to simply replacing $\matM$ with $\matM'$, which consists of $\ell$ copies of $\matM$ stacked on top of each other. 
\end{remark}

\begin{remark}If $\matM$ has column rank less than $m$, then the DLog problem is intuitively at least as hard as GMP. This is because we can use the DLog attack to recover $\vecv^{(b)}$, and then check if $\vecv^{(b)}$ is in the column-span of $\matM$. This reduction requires the DLog adversary to have success probability close to 1, which is why this relationship is not formal. Note that given many copies of each state, we use the random self-reducibility of DLog to boost the probability to be close to 1.
\end{remark}

\subsection{GMP Examples}

Here, we discuss how the GMP assumption encompasses many group action assumptions used in the literature.

\paragraph{DDH.} The DDH assumption on group actions states that $(a*x,b*x,(a+b)*x)$ is computationally indistinguishable from three random set elements. This can be framed as an instance of GMP by setting \[\matM=\left(\begin{array}{cc}1&0\\0&1\\1&1\end{array}\right)\;\;\;,\;\;\;\vecs=\left(\begin{array}{c}a\\b\end{array}\right)\]

\paragraph{LHS.} The Linear Hidden Shift (LHS) assumption~\cite{AC:ADMP20} hands out polynomially-many sample, which either have the form $(x_i,\vecv_i,(\vecv_i\cdot\vecs)*x_i)$ or $(x_i,\vecv_i,u_i)$, where $u_i$ is a random set element, $\vecs$ is a random vector in $\{0,1\}^n$, and $\vecv_i$ is a random vector in $\G^n$. The goal is to distinguish these two cases.

If the $x_i$ are actually all the same, and if the number of samples is a priori bounded by $m$, the LHS assumption is a instance of the GMP assumption using
\[\matM=\left(\begin{array}{c}\vecv_1^T\\\vdots\\\vecv_m^T\end{array}\right)\]

We can also handle the case where the $x_i$ are chosen independently (as in~\cite{AC:ADMP20}). For example, using the generalization of GMP to $\matM,\vecs$ containing both group elements and integers (see Remark~\ref{rem:mixmatch}), we can set
\[\matM'=\left(\begin{array}{cc}\mathbf{0}&\matI_m\\
\matM&\matI_m\end{array}\right)\;\;\;,\;\;\;\vecs'=\left(\begin{array}{c}\vecs\\\vect\end{array}\right)\]
where $\matI_m$ is the $m\times m$ identity matrix, and $\vect$ is a random vector of random group elements. Then the output of the GMP assumption is $2m$ set elements, and we let $x_i$ be the $i$th element, $u_i$ be the $(i+m)$th element.

\paragraph{Extended LHS.} In~\cite{TCC:AlaMalRah22}, a generalization of LHS is considered, called the extended LHS assumption. This assumption gives out several samples of the form $(\matM,\vecm,\vecx_0,\vecx_1,\vecy_0,\vecy_1)$, where: $\matM\in\G^{m\times m}$ is random, $\vecm\in\G^m$ is random, $\vecx_0,\vecy_0$ are independent vector of $m$ random set elements, and finally either $\vecx_1=(\matM\cdot\vecs)*\vecx_0$ and $\vecy_1=(\matM\cdot\vecs+\vecm\odot\vecs)*\vecy_0$, or $\vecx_1,\vecy_1$ are random.

If we let ${\sf Diag}(\vecm)$ denote the diagonal matrix whose elements are $\vecm$, then we can obtain the 1-sample extended LHS assumption by setting:

\[\matM'=\left(\begin{array}{ccc}\mathbf{0}&\matI_m&\mathbf{0}\\
\mathbf{0}&\mathbf{0}&\matI_m\\
\matM&\matI_m&\mathbf{0}\\
\matM+{\sf Diag}(\vecm)&\mathbf{0}&\matI_m\end{array}\right)\;\;\;,\;\;\;\vecs'=\left(\begin{array}{c}\vecs\\\vect_0\\\vect_1\end{array}\right)\]
Above, $\vect_0,\vect_1$ are vectors of random group elements. We can also obtain the extended LHS assumption for a bounded number of samples analogously.

%% file: coset.tex
\section{Coset Sampling on Quantum Group Actions}\label{sec:coset}

Here we explore how to adapt coset sampling attacks that apply in the case of classical group actions to the quantum state group action setting. Our main theorem is the following, which shows that coset sampling attacks apply when the group action is \emph{orthogonal}.

\begin{theorem}\label{thm:coset}For any \emph{orthogonal} and \emph{junk-free} quantum group action, there exists a polynomial $\ell$ and an \emph{exponential-time} quantum algorithm $\calA$ that solves the $\ell$-DLog problem with probability $1-\negl(\lambda)$. $\calA$ only makes $O(\ell)$ calls each to $\start$ and $\act$.
\end{theorem}
\begin{proof} In the classical group action setting, the analogous result follows from algorithms for the hidden sub-group problem in non-abeliang groups. Specifically, \cite{EttHoy00,EttHoyKni04} show an exponential time algorithm for the hidden subgroup problem, which in turn can be used to solve the discrete logarithm problem in abelian groups. Our observation is that these algorithms can be made to work, with a bit of extra work, on orthogonal junk-free quantum group actions as well.

Let $\G$ be a group and $\HH$ an unknown subgroup of $\G$. For a group element $g\in\G$, Define $|g\HH\rangle=\frac{1}{\sqrt{|\HH|}}\sum_{h\in\HH}|gh\rangle$. The states $|g\HH\rangle$ are called coset states. We will use the following lemma, implicit in~\cite{EttHoyKni04}:
\begin{lemma}\label{lem:coset} There exists a quantum algorithm $\calA$ which takes as input $\ell$ coset states $|g_1\HH\rangle,\cdots|g_\ell\HH\rangle$ for arbitrary $g_i\in\G$, and with probability $1-O(1/|\G|)$, outputs a set of generators for $\HH$. Here, $\ell=O(\log(|\G|)^4)$, and the running time of $\calA$ is $\poly(|\G|)$.
\end{lemma}

We now give our algorithm for solving DLog in quantum state group actions. We show the case of regular group actions, but the algorithm readily extends to irregular group actions as well with a more tedious analysis. Let $e$ be the identity in $\G$. Recall our convention that for regular quantum group actions, the states of the group action are labelled by the group elements themselves, with $\start$ producing $|\psi_e\rangle$, and $g*|\psi_h\rangle=|\psi_{gh}\rangle$.

We are given $\ell$ copies of $|\psi^{(1)}\rangle=g*|\psi_e\rangle$, for an $\ell$ to be determined below. Using $\start$, we also have $\ell$ copies of $|\psi^{(0)}\rangle:=|\psi_e\rangle$. To compute $g$,
\begin{enumerate}
    \item Repeat the following $\ell$ times:
    \begin{enumerate}
        \item Create the uniform superposition $\frac{1}{\sqrt{2}|\G|}\sum_{h_0,h_1\in\G,b\in\{0,1\}}|b,h_0,h_1\rangle$. Append one copy each of the states $|\psi^{(0)}\rangle$ and $|\psi^{(1)}\rangle$. Then the state is \[|\tau_a\rangle=\frac{1}{\sqrt{2}|\G|}\sum_{h_0,h_1\in\G,b\in\{0,1\}}|b,h_0,h_1\rangle|\psi^{(0)}\rangle|\psi^{(1)}\rangle\]
        \item Conditioned on $b=1$, swap the final two registers, obtaining
        \[|\tau_b\rangle=\frac{1}{\sqrt{2}|\G|}\sum_{h_0,h_1\in\G,b\in\{0,1\}}|b,h_0,h_1\rangle|\psi^{(b)}\rangle|\psi^{(1-b)}\rangle\]
        \item Apply the group action in superposition to obtain
    \begin{align*}
    |\tau_c\rangle&=\frac{1}{\sqrt{2}|\G|}\sum_{h_0,h_1\in\G,b\in\{0,1\}}|b,h_0,h_1\rangle(h_0*|\psi^{(b)}\rangle)(h_1*|\psi^{(1-b)}\rangle)\rangle\\
    &=\frac{1}{\sqrt{2}|\G|}\sum_{h_0,h_1\in\G,b\in\{0,1\}}|b,h_0,h_1\rangle|\psi_{h_0g^b}\rangle|\psi_{h_1g^{1-b}}\rangle\\
    &=\frac{1}{\sqrt{2}|\G|}\sum_{y_0,y_1\in\G,b\in\{0,1\}}|b, y_0g^{-b},y_1 g^{b}\rangle|\psi_{y_0}\rangle|\psi_{y_1g}\rangle
    \end{align*}
    where in the last line we used the substitution $h_0= y_0g^{-b}$ and $h_1=y_1g^{b}$.
        \item Discard the last two registers. Since the terms $|\psi_y\rangle$ for different $y$ are orthogonal, discarding these registers is equivalent to measuring $y_0$ and $y_1g$ (though the results of the measurement are hidden). This means the first three registers collapse to the state
        \[|\phi_{y_0,y_1}\rangle:=\frac{1}{\sqrt{2}}\left(|0,y_0,y_1\rangle+|1,y_0g^{-1},y_1g\rangle\right)\]
        for a random sample of $y_0,y_1$.
    \end{enumerate}
    \item Define a group $\G'=\G^2\rtimes\Z_2$ where and $\rtimes$ denotes the semi-direct product. In other words, $\G'$ comprises tuples $(b,g_0,g_1)\in\{0,1\}\times \G\times\G$, where the group operation is $(b,g_0,g_1)\cdot (c,h_0,h_1)=(b\oplus c,g_0 h_0^{(-1)^b},g_1 h_1^{(-1)^b})$. Then let $\HH$ be the subgroup generated by the element $(1,g^{-1},g)$, which has order 2. Thus, we see that $|\phi_{y_0,y_1}\rangle=|(0,y_0,y_1)\HH\rangle$, a coset state of $\HH$.
    \item We then feed these $\ell$ coset states into the algorithm from Lemma~\ref{lem:coset}. The result is the generator $(1,g,g^{-1})$ for $\HH$, which reveals $g$. In order to run Lemma~\ref{lem:coset}, we choose $\ell=O(\log(|\G'|)^4)=O(\log(|\G|)^4)$.\qed
\end{enumerate}
\end{proof}

\subsection{Other Coset Sampling Attacks}

\paragraph{Kuperberg's algorithm.} Kuperberg~\cite{Kuperberg05}, solves the dihedral hidden subgroup problem in sub-exponential time. This algorithm is readily adapted to give solve discrete logs for abelian group actions.

We can extend the algorithm to the case of quantum state group actions. One issue is that the algorithm requires a sub-exponential number of coset samples, which translates to needing a sub-exponential number of discrete log samples. In turn, since the correctness of a quantum state group acitons allows sample to incur a negligible error, the overall error of the sub-exponentially-many samples may be too high to ensure correctness. But the algorithm can be made to work by restricting to perfectly correct schemes. A straightforward adaptation of the techniques in Theorem~\ref{thm:coset} to Kuperberg's algorithm gives the following:

\begin{theorem}[Adaptation of~\cite{Kuperberg05}] For any orthogonal, junk-free, and \emph{perfectly} correct abelian quantum state group action, there exists a quantum algorithm $\calA$ running in time $2^{O(\sqrt{\log|\G|})}$ that breaks the $2^{O(\sqrt{\log|\G|})}$-DLog problem with probability $1-\negl(\lambda)$.
\end{theorem}

\paragraph{Simon's Algorithm.} Simon's algorithm\cite{Simon97} can be applied to group actions where the group is $\G=\Z_2^n$ to solve discrete logarithms in polynomial time. In particular, given a set element $y=g*x$, one can define $f(0,h)=h*x$ and $f(1,h)=h*y$. Then $f(b,h)$ has a period $(1,g)$, since $f((b,h)+(1,g))=f(b,h)$. Applying Simon's algorithm to $f$ then recovers $g$. By adapting the techniques from above to Simon's algorithm gives the following:

\begin{theorem}[Adaptation of~\cite{Simon97}] For any orthogonal, junk-free quantum state group action with group $\G=\Z_2^n$, there exists a polynomial $\ell$ and a quantum algorithm $\calA$ running in polynomial time such that $\calA$ breaks the $\ell$-DLog problem with probability $1-\negl(\lambda)$.
\end{theorem}

%% file: hashbased.tex
\section{A Hash-based Quantum State Group Action}\label{sec:hashbased}

\begin{construction}\label{constr:hashedbased}Let $\calR=(\calR_\lambda)_\lambda$ be a set. Let $N=N(\lambda)$ be an integer function, $H=(H_\lambda)_\lambda$ be a family of efficiently computable functions $H_\lambda:\calR_\lambda\rightarrow\Z_{N(\lambda)}$, and $(|\phi_\lambda\rangle)_\lambda$ be a family of efficiently constructible states where $|\phi_\lambda\rangle$ are superpositions over elements in $\calR_\lambda$. Let $(\start,\act)$ be the following:
\begin{itemize}
    \item $\start(1^\lambda):$ create the state $|\phi_\lambda\rangle$
    \item $\act(g,|\psi\rangle)$: $g$ is assumed to be an element in $\Z_{N(\lambda)}$. Let ${\sf P}_g$ be the unitary gate which maps $|r\rangle$ to $e^{i2\pi H_\lambda(r)\cdot g}|r\rangle$. Then output ${\sf P}_g|\psi\rangle$.
\end{itemize}
\end{construction}
The underlying classical group action can be taken as $\G_\lambda=\calX_\lambda=\Z_{N(\lambda)}$, with $g*x = g+x$. The states $|\psi_x\rangle$ are taken to be ${\sf P}_x|\phi_\lambda\rangle$. The distinguished starting element is $|\psi\rangle$. The construction is correct since ${\sf P}_g{\sf P}_h={\sf P}_{g+h}$ (where we have used additive notation for the group operation). It is also junk-free and regular. 

\paragraph{Conditions for Orthogonality.} Since there are $N(\lambda)$ set elements, and the dimension of the space is at most $|\calR_\lambda|$, orthogonality requires at a minimum that $H$ is not compressing. This is not enough, and whether or not Construction~\ref{constr:hashedbased} is orthogonal will depend on the particulars of $H$ and also the state $\ket{\phi_\lambda}$. We now give a sufficient condition:

\begin{lemma}\label{lem:orthogonality}Consider the distribution $\calD_\lambda$ obtained by measuring $|\phi_\lambda\rangle$ in the computational basis and then applying $H_\lambda$. If there is a negligible function $\negl(\lambda)$ bounding the statistical distance between $\calD_\lambda$ and uniform in $Z_{N(\lambda)}$, then the quantum state group action given in Construction~\ref{constr:hashedbased} is orthogonal.
\end{lemma}
\begin{proof}
Define \begin{align*}
      \ket{\psi_{g}}&= \sum_{j}\alpha_j\omega_{N}^{H(j)\cdot g}\ket{j}= \sum_{i=0}^{N-1}\left(\sum_{j:H(j)=i}\alpha_j\omega_{N}^{i\cdot g}\ket{j}\right)\\
  \end{align*}
Then $|\psi_0\rangle=|\phi_\lambda\rangle$. Then let $p_i=\sum_{h:H(j)=i}|\alpha_i|^2$. We see that $p_i=\Pr[i\gets\calD_\lambda]$. 

For the moment, suppose that $p_i=1/N$ for all $i$, indicating that $\calD_\lambda$ is uniform and hence $\negl(\lambda)=0$. In this case, for $g\neq g'$,
  \begin{align*}
\braket{\psi_g|\psi_g'}
&= \left(\sum_j |\alpha_j|^2 \omega_N^{H(j)\cdot(g'-g)}\right)\\
&= \sum_i \left(\sum_{j:H(j)=i}|\alpha_j|^2\right)\omega_N^{i\cdot (g'-g)}\\
&=\frac{1}{N}\sum_{i} \omega_N^{i\cdot (g'-g)}=0
\end{align*}
Thus, in this case the $|\psi_g\rangle$ are exactly orhogonal.
  
Now consider a state $\ket{\psi_g}$ such that the $p_i$ are not uniform. We are still given that $\calD_\lambda$ is statistically close to uniform, meaning $\sum_i |p_i-1/N|\leq\negl(\lambda)$ for some negligible function $\negl$. In other words,
    \[
    \frac{1}{2}\sum_{i \in \Z_N}\left|\sum_{j:H(j)=i}|\alpha_j|^2-\frac{1}{N(\lambda)}\right|\leq \negl(\lambda)
    \]
    
    Let $\ket{\psi'_{g}}=\sum_i\left(\frac{1}{\sqrt{N p_i}}\sum_{i:H(j)=i}\alpha_j\omega_N^{i\cdot g}|j\rangle\right)$. We will show that the $||\ket{\psi_g}-\ket{\psi'_g}||_2$ is negligible.
    \begin{align*}
        ||\ket{\psi_g}-\ket{\psi_g'}||_2^2&= \sum_i\left(\sum_{j:H(j)=i}|\alpha_j|^2\left(1-\sqrt{\frac{1}{Np_i}}\right)^2\right)\\
        &=\sum_i \left(\sqrt{p_i}-\sqrt{\frac{1}{N}}\right)^2\\
        &=\sum_i \left(\sqrt{\frac{1}{N}+t_i}-\sqrt{\frac{1}{N}}\right)^2
    \end{align*}
    where $t_i:= p_i-\frac{1}{N}$. Observe that $\sum_i |t_i|=\negl(\lambda)$.
    
    If $t_i \geq 0$, then $\sqrt{\frac{1}{N}+t_i}\leq \sqrt{\frac{1}{N}}+\sqrt{t_i}$, 
    and if  $t_i<0$, then
    $\sqrt{\frac{1}{N}}=\sqrt{\frac{1}{N}}+\sqrt{-t_i}-\sqrt{-t_i}\leq \sqrt{\frac{1}{N}-(-t_i)}+\sqrt{-t_i}$. Thus,
\begin{align*}
   \sum_{i=0}^{N-1}\left(\sqrt{\frac{1}{N}+t_i}-\sqrt{\frac{1}{N}}\right)^2
   &=  \sum_{i:t_i \geq 0}\left(\sqrt{\frac{1}{N}+t_i}-\sqrt{\frac{1}{N}}\right)^2+\sum_{i:t_i < 0}\left(\sqrt{\frac{1}{N}}-\sqrt{\frac{1}{N}-(-t_i)}\right)^2\\
   &\leq \sum_{i:t_i\geq 0}\left(\sqrt{t_i}\right)^2+\sum_{i:t_i < 0}\left(\sqrt{(-t_i)}\right)^2\\
   &= \sum_{i=0}^{N-1}t_i= \negl(\lambda)
    \end{align*}

Thus, we use the states $|\psi_g'\rangle$ as the underlying states in our group action, which are orthogonal. We then have that $\start$ and $\act$ produce the states $|\psi_g\rangle$, which are negligibly-close to $|\psi_g'\rangle$, giving correctness.
 \end{proof}
\begin{lemma}
\label{lem:kwiseind-orth}
Let $H= (H_{\lambda})_{\lambda}$ be a family of $k\geq 2$-wise independent  compressing hash functions $H_{\lambda}: \calR_\lambda\rightarrow\Z_{N(\lambda)}$ such that $H_{{\lambda}_{\infty}}(\mathcal{R}_{\lambda})\geq c \log N$, where $c> 2$ . Then, the quantum state group action given in Construction \ref{constr:hashedbased} instantiated with $H_{\lambda}$ is orthogonal.
\end{lemma}
\begin{proof}
The function $H:\calR_\lambda\rightarrow\Z_{N(\lambda)}$ is a $k$ wise independent hash function. From now on, we will drop the subscript $\lambda$ from $N(\lambda)$.\\
 Let $\cal{X}$ denote the distribution of the domain of the function $H$, with the guarantee that $H_{\infty}(\mathcal{X})\geq 4 \log N$. Let $\textbf{H}$ be the random variable distributed according to $\cal{H}$ and let its instantiation be $H$. The leftover hash lemma states that, $d_{\text{TV}}((\textbf{H},\textbf{H}(\cal{X})), (\textbf{H}, U_{\Z_{N}}))\leq \epsilon $, where $\epsilon=\frac{\sqrt{N}}{2^{4\log N}}$. More precisely, 
 \begin{align*}
    &\frac{1}{2}\sum_{z \in \Z_N, H\in \{0,1\}^d}\Big| \Pr[\textbf{H}(X)=z, \textbf{H}=H]-\Pr[U_{N}=z, U_{2^d}=H]\Big|\\
    &= \frac{1}{2}\sum_{z \in \Z_N, H\in \{0,1\}^d}\Big| \Pr[\textbf{H}(X)=z| \textbf{H}=H]\Pr[\textbf{H}=H]-\Pr[U_{N}=z]\cdot \Pr[ U_{2^d}=H]\Big|\\
    &= \frac{1}{2}\sum_{ H\in \{0,1\}^d}\Pr[\textbf{H}=H]\Big| \Pr[\textbf{H}(X)=z| \textbf{H}=H]-\Pr[U_{N}=z]\Big|\\
    &=\frac{1}{2}\sum_{ H\in \{0,1\}^d}\Pr[\textbf{H}=H]\Big| \Pr[H(X)=z]-\Pr[U_{N}=z]\Big|
 \end{align*}
 Observe that, at least $1-\sqrt{\epsilon}$ fraction of hash functions drawn from $\cal{H}$ must be such that, 
\begin{align*} 
\Big| \Pr[H(X)=z]-\Pr[U_{N}=z]\Big|&\leq \sqrt{\epsilon}\\
&\leq \frac{N^{1/4}}{2^{2\log N}}\\
&= \frac{N^{3/4}}{2^{2\log N} }\cdot \frac{1}{\sqrt{N}}\\
&= \frac{2^{0.75\log N}}{2^{2\log N} }\cdot \frac{1}{\sqrt{N}}\\
&= \negl(\lambda)\cdot \frac{1}{\sqrt{N}}
\end{align*}
This is because, if more than an $\epsilon$ fraction of hash functions drawn from $\mathcal{H}$ are such that $\Big| \Pr[H(X)=z]-\Pr[U_{N}=z]\Big|>\epsilon$, then $d_{\text{TV}}((\textbf{H},\textbf{H}(\mathcal{X})), (\textbf{H}, U_{\Z_{N}}))=\frac{1}{2}\sum_{ H\in \{0,1\}^d}\Pr[\textbf{H}=H]\Big| \Pr[H(X)=z]-\Pr[U_{N}=z]\Big|>\sqrt{\epsilon}\sqrt{\epsilon}=\epsilon$, a contradiction.\\
 Therefore, for at least $1-\sqrt{\epsilon}$ fraction of hash functions $H$ in the support of $\cal{H}$, $\Big|\Pr[H(x)=z]-\frac{1}{N}\Big|\leq \sqrt{\epsilon}.$ So the group action in Construction \ref{expandingcons} is orthogonal with probability at least $1-\frac{N^{1/4}}{2^{2\log N}}$ over the choice of $H$.
 \end{proof}

%% file: generalized_matrix_assumption.tex
\section{Generalized Matrix Assumption}\label{sec:gmp1}

In this section, we will prove that the generalized matrix problem is hard in each of the following cases:
\begin{itemize}
\item $H$ is a $2n$ wise independent expanding function. Here $n$ denotes the number of rows of the matrix.
\item $H$ is the composition of a $2n$ wise independent hash function with a function sampled from the injective mode of a lossy function family. 
\item $H$ is the random oracle.   \end{itemize}

Before we prove the main theorem, we state and prove the following lemma:
\begin{lemma}
\label{lem:densmatrix}
    Fix a security parameter $\lambda$ and an instance $(\matM,\vecv^{(b)}*\ket{\psi_{x_{\lambda}}}^n)$ of the $(\cal{M},\cal{S})-$Generalized Matrix Problem in a hash based quantum state group action from Construction \ref{constr:hashedbased} instantiated with the hash function $H_{\lambda}: \mathcal{R_{\lambda}}\rightarrow \Z_{N(\lambda)}$. Let $j$ be the number of distinct rows in $\matM$. For all $i \in [j]$, let $S_i$ be the set of all row indices equal to the $i^{th}$ distinct row. Let  $\matM'\in \Z_{N}^{j \times m}$ be the same matrix as $\matM$ but without repeated rows. Then,
    \begin{enumerate}
        \item If $b=0$: The resulting mixed state is equivalent to the density matrix obtained by the following procedure:
        \begin{itemize}
        \item Initialize quantum registers $X_1, \dots X_n$  and $Y_1, \dots Y_j$  to states $\otimes_{i =1}^{n}\ket{0}_{X_i}\otimes_{i =1}^{j}\ket{0}_{Y_i}$.
        \item Prepare the state  $\frac{1}{\sqrt{|R_{\lambda}|^n}}\sum_{x_1, \dots x_n\in R_{\lambda}^n}\ket{x_1}_{X_1} \dots \ket{x_n}_{X_n}\otimes_{i =1}^{j}\ket{0}_{Y_i}$. 
            \item Apply in superposition the map $(\otimes_{k \in S_i}\ket{x_k}_{X_k})\ket{0}_{Y_i}\rightarrow (\otimes_{k \in S_i}\ket{x_k}_{X_k})\ket{\sum_{k \in S_i}H_{\lambda}(x_k)}_{Y_i}$ , yielding the state $\frac{1}{\sqrt{|R_{\lambda}|^n}}\sum_{x_1, \dots x_n}\otimes_{i=1}^{n}\ket{x_i}_{X_i} \otimes_{i \in [j]}\ket{\sum_{k \in S_i}H_{\lambda}(x_k)}_{Y_i}$.
            \item Measure the registers $Y_1, Y_2, \dots Y_j$, giving $(y_1, \dots y_j)$, collapsing the state on registers $X_1, \dots X_n$ to $\sum_{\substack{x_1, \dots x_n:\\\forall i \in [j]: \sum_{k \in S_i}H(x_k)=y_i}}\otimes _{i=1}^{n}\ket{x_i}_{X_i}$.
        \end{itemize}
        
        \item If $b=1$:  Then the resulting mixed state is equivalent to the density matrix obtained by the following procedure:
        \begin{itemize}
       \item Initialize quantum registers $X_1, \dots X_n$  and $Y_1, \dots Y_j$  to states $\otimes_{i =1}^{n}\ket{0}_{X_i}\otimes_{i =1}^{j}\ket{0}_{Y_i}$.
        \item Prepare the state  $\frac{1}{\sqrt{|R_{\lambda}|^n}}\sum_{x_1, \dots x_n\in R_{\lambda}^n}\ket{x_1}_{X_1} \dots \ket{x_n}_{X_n}\otimes_{i =1}^{j}\ket{0}_{Y_i}$. 
            \item Apply in superposition the map $(\otimes_{k \in S_i}\ket{x_k}_{X_k})\ket{0}_{Y_i}\rightarrow (\otimes_{k \in S_i}\ket{x_k}_{X_k})\ket{\sum_{k \in S_i}H_{\lambda}(x_k)}_{Y_i}$ , yielding the state $\frac{1}{\sqrt{|R_{\lambda}|^n}}\sum_{x_1, \dots x_n}\otimes_{i=1}^{n}\ket{x_i}_{X_i} \otimes_{i \in [j]}\ket{\sum_{k \in S_i}H_{\lambda}(x_k)}_{Y_i}$.
            \item Measure the state on $\matM'^T\cdot  (\sum_{k \in S_1}H(x_k), \dots,\sum_{k \in S_j}H(x_k) )$, giving $(y_1, \dots y_j)$ and  collapsing the state to\\ $\sum_{\substack{x_1, \dots x_n:\\ \matM'^T\cdot  (\sum_{k \in S_1}H(x_k), \dots,\sum_{k \in S_j}H(x_k) )= (y_1, \dots y_j)}}\otimes _{i=1}^{n}\ket{x_i}_{X_i}$.
        \end{itemize}
    \end{enumerate}
\end{lemma}

\begin{proof}
First, some notation: let $x$ denote $x_1, \dots x_n$, and let $x_1'$ denote $x_1', \dots x_n'$.\\
    We will analyse the case when $b=0$. In this case, $\cal{A}$ receives $\{\ket{\psi}_{v^{(0)}_1},\ket{\psi}_{v^{(0)}_2}, \dots , \ket{\psi}_{v^{(0)}_n}\} $, where $v^{(0)}_i$ is the $i^{th}$ coordinate of $v^{(0)}$. Let $|{\sf EqPat}_\matM(\Z_N^n)|=S$.  The density matrix representing the corresponding mixed state is:  
\begin{align*}
&\frac{1}{S}\sum_{v^{(0)}_1, \dots v^{(0)}_n}\ket{\psi}_{v^{(0)}_1}\bra{\psi}_{v^{(0)}_1}\otimes\dots \otimes  \ket{\psi}_{v^{(0)}_n}\bra{\psi}_{v^{(0)}_n}\\
&=\frac{1}{S2^{kn}}\sum_{\substack{v^{(0)}_1, \dots v^{(0)}_n,\\x_1,x_1',\dots x_n,x_n'}}\omega_N^{H(x_1)-H(x_1')\cdot v^{(0)}_1 }\ket{x_1}\bra{x_1'}\otimes \dots \otimes \omega_N^{H(x_n)-H(x_n')\cdot v^{(0)}_n }\ket{x_n}\bra{x_n'}\\
&=\frac{1}{S2^{kn}}\sum_{x_1,x_1'}\sum_{v^{(0)}_1}\omega_N^{H(x_1)-H(x_1')\cdot v^{(0)}_1 }\ket{x_1}\bra{x_1'}\otimes \dots \otimes \sum_{x_n,x_n'}\sum_{v^{(0)}_n}\omega_N^{H(x_n)-H(x_n')\cdot v^{(0)}_n }\ket{x_n}\bra{x_n'}\\
\end{align*}
Let $j\leq n$ be the number of distinct elements in $v^{(0)}$, and WLOG, let these $j$ elements be $v^{(0)}_{S_1}, \dots v^{(0)}_{S_j}$. Now, observe that we can partition the coordinates of $v^{(0)}$ into sets $S_i: i \in \{1, \dots ,j\} $, where the set $S_i=\{k \in \{1, \dots n\}: v_k^{(0)}= v^{(0)}_{S_i}\}$. Using this partition, we can further simplify the last expression as follows:
    \begin{align*}
        &\frac{1}{S2^{kn}}\sum_{x_1,x_1'}\sum_{v^{(0)}_1}\omega_N^{H(x_1)-H(x_1')\cdot v^{(0)}_1 }\ket{x_1}\bra{x_1'}\otimes \dots \otimes \sum_{x_n,x_n'}\sum_{v^{(0)}_n}\omega_N^{H(x_n)-H(x_n')\cdot v^{(0)}_n }\ket{x_n}\bra{x_n'}\\
        &= \frac{1}{S2^{kn}}\sum_{x,x'}\sum_{v^{(0)}_{S_1}}\omega_N^{v^{(0)}_{S_1}\sum_{i \in S_1}H(x_i)-H(x_i')}\dots\sum_{v^{(0)}_{S_j}}\omega_N^{v^{(0)}_{S_j}\sum_{i \in S_j}H(x_i)-H(x_i')}\ket{x_1}\bra{x_1'}\dots \ket{x_n}\bra{x_n'}\\
        &=  \frac{1}{2^{kn}}\sum_{x,x'\text{ s.t. }\forall k \in \{1, \dots j\}: \sum_{i \in S_k}H(x_i)-H(x_i')=0}\ket{x_1}\bra{x_1'}\dots \ket{x_n}\bra{x_n'}\\
        \end{align*}
 where $\omega_N=e^{2\pi i/N}$, and the second equality follows because $ \sum_{g\in\mathbb{G}}\omega^{g\cdot i}=0 $ whenever $i\neq 0$.\\
 Next, we will  analyse the case when $b=1$.  In this case, $\cal{A}$ receives $\{\ket{\psi}_{v^{(1)}_1},\ket{\psi}_{v^{(1)}_2}, \dots , \ket{\psi}_{v^{(1)}_n}\} $, where $v^{(1)}_i=(\matM\cdot\vecs)_i$. The density matrix representing the corresponding mixed state is:  

 \begin{align*}
    &\sum_{\vecs}\Pr[\vecs](\matM\cdot \vecs)_1*\ket{\psi}_0\bra{\psi}_0\otimes (\matM\cdot \vecs)_2*\ket{\psi}_0\bra{\psi}_0\dots \otimes (\matM\cdot \vecs)_n*\ket{\psi}_0\bra{\psi}_0\\
    &=\frac{1}{2^{kn}}\sum_{\vecs,x_1,x_1'\dots x_n,x_n'}\Pr[\vecs]\omega_{N}^{(H(x_1)-H(x'_1))\cdot\langle \matM_1, \vecs\rangle} \ket{x_1}\bra{x_1'}\otimes \dots \otimes \omega_{N}^{(H(x_n)-H(x'_n))\cdot \langle \matM_n, \vecs\rangle} \ket{x_n}\bra{x_n'}\\
     \end{align*}

    Let $j\leq n$ be the number of distinct rows in $\matM$. Now, observe that we can partition the rows of $\matM$ into sets $S_i: i \in \{1, \dots ,j\} $, where the set $S_i$ contains all rows that are equal to the $i^{th}$ distinct row. Let $\matM'\in \Z_N^{j \times m}$ be the same matrix as $\matM$ but without any colliding rows. \\
    Using this partition, we can further simplify the last expression as follows:
    \begin{align*}
        &\frac{1}{ 2^{kn}}\sum_{\vecs,x,x'}\Pr[\vecs]\omega_{N}^{(H(x_1)-H(x'_1))\langle \matM_1, \vecs \rangle} \ket{x_1}\bra{x_1'}\otimes \dots \otimes \omega_{N}^{(H(x_n)-H(x'_n))\langle \matM_n, \vecs\rangle} \ket{x_n}\bra{x_n'}\\
        &= \frac{1}{2^{kn}}\sum_{\substack{x,x'}}\sum_{\vecs}\Pr[\vecs]\omega_N^{\langle \matM'_1,\vecs\rangle \sum_{i \in S_1}(H(x_i)-H(x_i'))}\dots\omega_{N}^{\langle \matM'_j, \vecs\rangle\sum_{i \in S_j}(H(x_i)-H(x_i'))}\ket{x_1}\bra{x_1'}\dots \ket{x_n}\bra{x_n'}\\
        \end{align*}
        For simplicity of notation,  let $\rho= (\sum_{i \in S_1}H(x_i), \dots \sum_{i \in S_j}H(x_i))$ and let $\rho'= (\sum_{i \in S_1}H(x_i'), \dots \sum_{i \in S_j}H(x_i'))$. Then,
        \begin{align*}
            &\frac{1}{2^{kn}}\sum_{\substack{x,x'}}\sum_{\vecs}\Pr[\vecs]\omega_N^{\langle \matM'_1,\vecs\rangle \cdot (\rho-\rho)_1}\dots\omega_{N}^{\langle \matM'_j, \vecs\rangle(\rho-\rho)_j}\ket{x_1}\bra{x_1'}\dots \ket{x_n}\bra{x_n'}\\
            &=    \frac{1}{2^{kn}}\sum_{\substack{x,x'}}\sum_{\vecs}\Pr[\vecs]\omega_N^{\langle (\rho-\rho)_1\cdot\matM'_1,\vecs\rangle }\dots\omega_{N}^{\langle (\rho-\rho)_j\cdot\matM'_j,\vecs\rangle}\ket{x_1}\bra{x_1'}\dots \ket{x_n}\bra{x_n'}\\
            &=    \frac{1}{2^{kn}}\sum_{\substack{x,x'}}\sum_{\vecs}\Pr[\vecs]\omega_N^{\langle \matM'^T_1\cdot(\rho-\rho)_1,\vecs\rangle }\dots\omega_{N}^{\langle \matM'^T_j\cdot(\rho-\rho)_j,\vecs\rangle}\ket{x_1}\bra{x_1'}\dots \ket{x_n}\bra{x_n'}\\
            &=\frac{1}{ 2^{kn}}\sum_{\substack{x,x'}}\sum_{\substack{\vecs_1\\\dots \vecs_m}}\Pr[\vecs]\omega_N^{\vecs_1\sum_{i=1}^{j}\matM'_{i1}(\rho-\rho')_i}\dots \omega_N^{\vecs_m\sum_{i=1}^{j}\matM'_{im}(\rho-\rho')_i}\ket{x_1}\bra{x_1'}\dots \ket{x_n}\bra{x_n'}
        \end{align*}
       
       For $t \in [m]$, observe that if $\sum_{i=1}^{j}\matM'_{it}(\rho-\rho')_i\neq 0$, then \\$\sum_{\vecs_t}\omega_N^{\vecs_t\sum_{i=1}^{j}\matM'_{it}(\rho-\rho')_i}=0$. \\
         
       So we can rewrite the final state as follows:
       \[
       \frac{1}{ 2^{kn}}\sum_{\substack{x_1,x_1',\dots,  x_n,x_n': \\  \rho\cdot \matM' =\rho'\cdot  \matM'  }}\ket{x_1}\bra{x_1'}\dots \ket{x_n}\bra{x_n'}
       \]
      \qed \end{proof}
          Now, for an adversary $\cal{A}$  with access to $H$ consider the following security game: 
\paragraph{Expt$^{\cal{A}}(1^{\lambda},b)$}
\begin{enumerate}
    \item The challenger creates the state \[\frac{1}{\sqrt{|R_{\lambda}|^n}}\sum_{x_1, \dots x_n}\otimes_{i=1}^{n}\ket{x_i}_{X_i} \otimes_{i \in [j]}\ket{\sum_{k \in S_i}H_{\lambda}(x_k)}_{Y_i}\].
    \item If $b=0$, the challenger measure the state on $\rho=(\sum_{k \in S_1}H(x_k), \dots \sum_{k \in S_j}H(x_k))$, and sends the resulting state to $\calA$. If $b=1$, it measures the state on $ \rho\cdot M'$, and sends the resulting state to $\calA$.
    \item  $\cal{A}$  returns a bit $b'$.
\end{enumerate}
 From now on, we will denote the above security game by \textbf{G}.
We say that \textbf{G} is secure if the following holds for all $\mu\in \{0,1\}$
\[
\Big| \Pr[\mu\leftarrow Expt^{\cal{A}}(1^{\lambda},0)]-\Pr[\mu\leftarrow Expt^{\cal{A}}(1^{\lambda},1)]\Big|\leq \negl(\lambda)
\]

In the following sections, we will present constructions of hash functions such that $\textbf{G}$ is secure. Observe that from Lemma \ref{lem:densmatrix}, this proves that the generalized matrix problem is hard in hash based quantum group actions instantiated with hash functions from these constructions.  

%% file: k-wise_ind_expanding_H.tex
\subsection{Expanding $2n$-wise independent $H$}
Consider a family of hash functions $H= \{H_{\lambda}\}_{\lambda}$ with the following construction:
\begin{construction} \label{expandingcons} Let $k,n, N$ be polynomials in $\lambda$ and let  $N$ be a prime such that $n,k \leq \frac{\sqrt{\log N}}{2}$. 
Let $\{0,1\}^k$ be the domain of $H_{\lambda}$. Set $H_{\lambda}$ to be sampled from a uniform and $2n$ wise independent function distribution $\mathcal{H}_{\lambda}:\{0,1\}^k\rightarrow \Z_N$. Since $\frac{2^k}{|\Z_N|}\leq \frac{2^k}{2^{4k}}= \negl(\lambda)$, $H_{\lambda}$ is injective with overwhelming probability on any $2n$ wise independent function distribution.
\end{construction}
\begin{remark}
    Since $H_{\lambda}$ is expanding, $H_{\infty}(\mathcal{X})\leq k\leq \frac{\sqrt{\log N}}{2}$. This implies that a quantum group action instantiated with $H_{\lambda}$ won't be orthogonal. However, this is to be expected because, as we will show later in this section, the generalized matrix problem instantiated with $2n$ wise independent and expanding hash based group actions is hard even against computationally unbounded adversaries.
\end{remark}

\noindent Now, we will prove the following lemma:
\begin{lemma}
\label{lem:MH-inj}
   Let $j,m,n$ be polynomials in $\lambda$ such that $j \leq m$. Let $\matM \in \Z_{N}^{j\times m}$ be such that, $\forall r\neq s \in [j], \matM_r \neq \matM_s$, and $\forall i \in [j]: \matM_i \neq 0$. Let $x=(x_1, \dots x_n)$ and $x'=(x_1', \dots, x_n')$ be disjointly paritioned into sets $S_1, \dots S_j$.  Then, with probability at least $1- \text{negl}(\lambda)$ over the choice of $H_{\lambda}$, \\$\matM^T\cdot (\sum_{i\in S_1}H_{\lambda}(x_i),\sum_{i\in S_2}H_{\lambda}(x_i),\dots \sum_{i\in S_j}H_{\lambda}(x_i))$ is injective on  $(x_1, \dots x_n)$, upto reordering $x_k \in S_i, \forall i \in [j]$.
\end{lemma}

\begin{proof}Let $x=(x_1,\cdots,x_n)$ and $x'=(x_1',\cdots,x_n')$. We denote
\[\left(\sum_{i\in S_1}H(x_i),\dots \sum_{i\in S_j}H(x_i)\right)\text{ by }y\text{, and }
\left(\sum_{i\in S_1}H(x_i')\dots \sum_{i\in S_j}H(x_i')\right)\text{ by }y'.
\]
We let  $H_{\lambda}$ be $H$. We denote the set of distinct $n$ tuples by $\text{dist}(\{0,1\}^{k},n)$.\\ We will first prove that, for all $x\in \text{dist}(\{0,1\}^{k},n), \matM^T\cdot y$ is injective on $x$. Then, we will prove that the weight of tuples $\notin \text{dist}(\{0,1\}^{k},n)$ is negligible. \\
First, consider the tuples $x,x'\in \text{dist}(\{0,1\}^k,n\})$. We will evaluate the probability that $\matM^T\cdot y'= \matM^T\cdot y$. Let $E$ be the event that $\matM^T\cdot y'= \matM^T\cdot y$. \\
We have the following cases:
\begin{enumerate}
    \item \textbf{Case 1:} There exists $x_i'\neq x_i$ such that $\forall j \in [n]: x_i\neq x_j',x_i'\neq x_j$. 
    \item \textbf{Case 2:} There exists $x_i \neq x_i'$ such that $\forall j \in [n]: x_i\neq x_j'$ or $\forall j \in [n]: x_i'\neq x_j$. 
\end{enumerate}
\begin{claim}
   Every pair of tuples $ x\neq x' \in \text{dist}(\{0,1\}^k,n)$ is either in Case 1 or 2.
\end{claim}
\begin{proof}
First, observe that since  $x\neq x'$, there must exist $i\in [n]$ such that $x_i \neq x_i'$. Now, let's assume for contradiction that for all pairs $x_i\neq x_i'$, $\exists j \in [n]: x_i= x_j'$ and $\exists k \in [n]: x_i'\neq x_k$. This implies that, for all $x\neq x':$  $\sum_{i}(H(x_i)-H(x_i'))=0$. Thus, for all $x\neq x'$ there exists a permutation $\pi$\footnote{When $x=x'$, the permutation is simply the identity.} such that 
\begin{align*}
    \matM^T\cdot y&=\matM^T\cdot(\sum_{i\in S_1}H(\pi(x_i)),\sum_{i\in S_2}H(\pi(x_i)),\dots \sum_{i\in S_j}H(\pi(x_i))) \\
    &=\pi(M^T)\cdot y
\end{align*}
 where $\pi$ permutes the columns of $M^T$. However, this implies that the corresponding columns of $M^T$ are identical, a contradiction.\qed\end{proof}

If $M^T\cdot y'= M^T\cdot y$, then $M^T\cdot (y-y')= 0$. Since every element in the vectors $(x_1, \dots x_n)$ and $(x_1', \dots x_n')$ is distinct, we can conclude that in Case 1,  $x_i$ and $x_i'$ can be chosen independently of all other elements. Also,  $\forall a \in \Z_N,  \Pr_H[H(x_i)=a]=\frac{1}{N}$  from the $2n$ wise independence of $H$. Without loss of generality, let $x_i, x_i'\in S_d$. Since all rows of $M$ are non-zero, there must exist a non-zero element in $M_d$. Now, because $\sum_{i\in S_d}(H(x_i)-H(x_i'))$ is uniformly and independently distributed over $\Z_N$, in this case we can conclude that $\Pr_H[M^T\cdot (y-y')= 0]\leq\frac{1}{N}$. \\
In case 2, one of the elements $x_i$ or $x_i'$ is equal to another element $x_j$ or $x'_k$ But in this case as well, since $H(x_i)$ (or $H(x_i')$) is uniform and $H(x_i)$ (or $H(x_i')$) is independent of all other elements, $\sum_{i\in S_d}(H(x_i)-H(x_i'))$ is uniformly distributed over $\Z_N$. So $\Pr_H[\matM^T\cdot (y-y')= 0]\leq \frac{1}{N}$.\\
So from the union bound, 
\begin{align*}
    &\Pr_H[\exists x\neq x' \in \text{dist}(\{0,1\}^k,n) : M^T\cdot (y-y')= 0 ]\\
    &\leq 2^{2kn}\Pr_H[ M^T\cdot( y-y')= 0 ]\\
&=2^{2kn}\frac{1}{N}
\leq \frac{2^{\log (N)/2}}{N}
= \negl(\lambda)
\end{align*}

\begin{align*}
    \Pr[E] &\leq \Pr[E|x,x'\in \text{dist}(\{0,1\}^k,n)]\cdot \Pr[\text{dist}(\{0,1\}^k,n)]+\\ &\qquad \Pr[E|x,x'\notin\text{dist}(\{0,1\}^k,n) ]\cdot \Pr[x,x'\notin\text{dist}(\{0,1\}^k,n)]\\
    &\leq \Pr[E|x,x'\in\text{dist}(\{0,1\}^k,n)]+ \Pr[x,x'\notin\text{dist}(\{0,1\}^k,n)]\\
    &\leq  \frac{2^{\log (N)/2}}{N} + (\max_{x_i}{W_{x_i}})^n\\
\end{align*}
where $W_{x_i}$ denotes probability of $x_i$ being sampled from the domain distribution. 
As long as $\max_{x_i}W_{x_i}$ is negligible, the probability of the event $E$ would be negligible and the statement of the lemma holds.\qed\end{proof}

We will now prove that the generalized matrix problem is secure. 
\begin{theorem}\label{thm:kwiseind}Fix a security parameter $\lambda$, and let $H_{\lambda}:\{0,1\}^k\rightarrow \Z_N$ be a hash function from Construction \ref{expandingcons}. Let $m,n$ be polynomials in $\lambda$, and let $N$ be a prime such that $k,n \leq \frac{\sqrt{\log N}}{2}$. Let $\calM_\lambda$ be a distribution over  $\Z_{N}^{n\times m}$ and let $\calS_\lambda$ be a distribution over $\Z_{N}^{m}$. Then, for all quantum adversaries $\cal{A}$, the $(\calM,\calS)$-Generalized Matrix (GM) Problem is hard in a quantum group action instantiated with $H_{\lambda}$. 
\end{theorem}
\begin{proof}
From Lemma \ref{lem:densmatrix}, it suffices to prove that the security game $\textbf{G}$ is secure. Formally, we will prove the following lemma:
\begin{lemma}
    Fix a security parameter $\lambda$. For any Adversary $\cal{A}$  with oracle access to $H_{\lambda}$ in $\textbf{G}$, , there exists a negligible function $\text{negl}$ such that
    \[
    \Pr[1\leftarrow Expt^{\cal{A}}(1^{\lambda},1)]-\Pr[1\leftarrow Expt^{\cal{A}}(1^{\lambda},0)]\leq \text{negl} (\lambda)
    \]
\end{lemma}
\begin{proof} We introduce the following sequence of hybrids:

    \noindent \textbf{Hybrid 0}:
    This is the same as $Expt^{\cal{A}}(1^{\lambda},1)$.\\
    \textbf{Hybrid 1}: This is the same as \textbf{Hybrid 0} except the state in $Expt^{\cal{A}}(1^{\lambda},1)$  is measured on $(\sum_{g \in S_1}x_g, \dots,\sum_{g \in S_j}x_g)$.\\
\begin{claim}

    For any  adversary $A$, there exists a negligible function $\lambda$ such that
    \[
    \Big|\Pr[A(H_0)=1]-\Pr[A(H_1)=1]\Big|\leq \text{negl}(\lambda) 
    \]
    
\end{claim}
\begin{proof}
Let $\mathcal{X}$ be the distribution over the domain of the function $H$. Because $\mathcal{X}$ is uniform, 
$\max_{x_i}W_{x_i}= \frac{1}{2^k}$. Then, 
\[
    \Big|\Pr[A(H_0)=1]-\Pr[A(H_1)=1]\Big|\leq \text{negl}(\lambda)\leq \frac{1}{2^{nk}}+\frac{1}{\sqrt{N}}
    \]
This follows directly from Lemma \ref{lem:MH-inj}.

\qed \end{proof}

\noindent\textbf{Hybrid 2}: This is the same as \textbf{Hybrid 1} except the state  \[\frac{1}{2^{kn/2}}\sum_{x_1, \dots x_n}\ket{x_1}\ket{H(x_1)}\dots \ket{x_n}\ket{H(x_n)}
\]is measured on $(\sum_{g \in S_1}H(x_g), \dots,\sum_{g \in S_j}H(x_g))$.\\
 \begin{claim}
    For any  adversary $A$, there exists a negligible function $\lambda$ such that
    \[
    \Big|\Pr[A({\bf Hybrid~2})=1]-\Pr[A({\bf Hybrid~1})=1]\Big|\leq \text{negl}(\lambda) 
    \]
    
\end{claim}
\begin{proof}
Let $x=x_1, \dots x_n$, $x'=x_1', \dots x_n', \rho=\sum_{g \in S_1}H(x_g), \dots,\sum_{g \in S_j}H(x_g), \rho'=\sum_{g \in S_1}H(x'_g), \dots,\sum_{g \in S_j}H(x'_g)$.
We prove that, for $\forall i \in [j]$, $\rho_i$, with high probability over $H$, $\rho_i$ is injective on $(x_1, \dots ,x_{|S_i|})$ up to reordering. 

 \begin{align*}
     \Pr[ \exists (x'_g)_{g\in S_i} \neq (x_g)_{g\in S_i}: \rho_i=\rho'_i]&\leq 2^{2ki}\cdot  \Pr\left[  \sum_{g\in S_i}H(x_g')-\sum_{g\in S_i}H(x_g)=0\right]\\
     &= \frac{2^{2ki}}{N}\
     \leq \frac{2^{2nk}}{N}
     = \frac{2^{\log N/2}}{N}
     = \frac{1}{\sqrt{N}}
 \end{align*}
 From the union bound,
 \begin{align*}
   \Pr[\exists x\neq x':\rho=\rho'] \leq  j\Pr[ \exists x_1', \dots x_i' \neq x_1, \dots x_i: \rho_i=\rho'_i]\leq 
   \frac{n}{\sqrt{N}}\tag*{\qed}
 \end{align*}
\end{proof}

 Thus, we have proven that, for any QPT adversary $\cal{A}$, $\textbf{G}$ is secure. Now Theorem \ref{thm:kwiseind} follows directly from Lemma \ref{lem:densmatrix}.
 \end{proof}

%% file: lossyH.tex
 \subsection{GMP with Lossy $H$}\label{lossyH}
Note that Construction \ref{expandingcons} is not orthogonal. In this section, we show how to use lossy functions to obtain an orthogonal quantum state group action for which the GMP assumption holds with computational security.

Before we state the construction of $\{H_{\lambda}\}_{\lambda}$, we will formally define lossy functions.
\begin{definition}[Lossy Functions]
    A $r$ lossy function is a keyed family of hash functions with two ways to sample the key $q \leftarrow \text{InjSamp}(1^{\lambda})$ and $q \leftarrow \text{LossySamp}(1^{\lambda})$. Each key $q$ defines hash function $f_q: \{0,1\}^{k} \rightarrow  \{0,1\}^{\ell}$, and has the following properties:
    \begin{enumerate}
        \item With overwhelming probability over $q \leftarrow \text{InjSamp}(1^{\lambda})$, $f_{q}$ is injective, meaning that $|\{y:\exists x \text{ s.t. }  f(x)=y\}|=2^{k}$.
        \item With overwhelming probability over $q \leftarrow \text{LossySamp}(1^{\lambda})$, $f_{k}$ is $r$ lossy, meaning that $|\{y:\exists x \text{ s.t. }  f(x)=y\}|\leq 2^{k-r}$.
        \item For any QPT adversary $\{\cal{A_{\lambda}}\}_{\lambda}$, 
        \[
        \Big|\Pr_{q \leftarrow \text{InjSamp}(1^{\lambda})}[\mathcal{A_{\lambda}}(q)\rightarrow 1]- \Pr_{q \leftarrow \text{LossySamp}(1^{\lambda})}[\mathcal{A_{\lambda}}(q)\rightarrow 1]\Big|\leq \negl(\lambda)
        \]

    \end{enumerate}
\end{definition}
Such lossy functions can be constructed from LWE~\cite{STOC:PeiWat08}, and we can take $k-r$ to be a fixed polynomial in $\lambda$~\cite{C:WatZha24}. We will  prove the  generalized matrix assumption in the case when $H$ has the following construction:
\begin{construction}
\label{lossycons}
 Let $k,n,r$ be polynomials in $\lambda$.
Let  $N$ be a prime such that $k \geq 8 \log N$, and  $2n(k-r)\leq \frac{\log N}{4}$. $H_{\lambda}: \{0,1\}^{k} \rightarrow \Z_N= h \circ f_q$ where $h:\{0,1\}^{\ell}\rightarrow \Z_N$ is a $2n$ wise independent hash function and $q\leftarrow \text{InjSamp}(1^{\lambda})$.
\end{construction}
\begin{lemma}
   The family of  functions $\{H_{\lambda}\}_{\lambda}$ in  Construction \ref{lossycons} yields an orthogonal group action.
\end{lemma}
\begin{proof}
We will first prove that, with overwhelming probability over the choice of $f_q$, the function $H=h\circ f_q$ in Construction \ref{lossycons} is $2n$ wise independent. For any $2n$ distinct elements $x_1, x_2, \dots x_{2n}\in \{0,1\}^k$, with overwhelming probability over $f_q$, we have that:
\begin{align*}
  \Pr_{H}[H(x_1)=y_1\wedge \dots   \wedge H(x_{2n})=y_{2n}]\\
  &= \Pr_{h,f_q}[h(f_q(x_1))=y_1\wedge \dots  \wedge h(f_q(x_{2n}))=y_{2n}]\\
  &=  \Pr_{h, z_1\neq z_2 , \dots \neq z_{2n}}[h(z_1)=y_1\wedge \dots   \wedge h(z_{2n})=y_{2n}]\\
  &= \frac{1}{N^{2n}}
  \end{align*} 
  where the second equality follows because $f_q$ is an injective function with overwhelming probability.\\
  Also observe that $H_{\infty}(\mathcal{X})=k\geq 8 \log N$.
  Now the proof follows directly from Lemma \ref{lem:kwiseind-orth}.
   
   \end{proof}

\end{proof}

\begin{theorem}\label{thm:lossy}Fix a security parameter $\lambda$, and let $H_{\lambda}:\{0,1\}^k\rightarrow \Z_N$ be a hash function from Construction \ref{lossycons}. Let $m,n$ be polynomials in $\lambda$, and let $N$ be a prime such that $2n(k-r) \leq \frac{\log N}{4}$. Let $\calM_\lambda$ be a distribution over  $\Z_{N}^{m\times n}$ and let $\calS_\lambda$ be a distribution over $\Z_{N}^{m}$. Then, for all QPT adversaries $\cal{A}$ the $(\calM,\calS)$-Generalized Matrix (GM) Problem is hard in a quantum group action instantiated with $H_{\lambda}$. 
\end{theorem}
\begin{proof}
From Lemma \ref{lem:densmatrix}, it suffices to prove that the security game $\textbf{G}$ is secure. Formally, we will prove the following lemma:
\begin{lemma}
  Fix a security parameter $\lambda$. For any QPT Adversary $\cal{A}$  with oracle access to $H_{\lambda}$ in $\textbf{G}$, , there exists a negligible function $\text{negl}$ such that
    \[
    \Pr[1\leftarrow Expt^{\cal{A}}(1^{\lambda},1)]-\Pr[1\leftarrow Expt^{\cal{A}}(1^{\lambda},0)]\leq \text{negl} (\lambda)
    \]
\end{lemma}
\begin{proof} We introduce the following sequence of hybrids:

    \noindent \textbf{Hybrid 0}: This is the same as  Expt$^{\cal{A}}(1^{\lambda},1)$.\\
    \noindent \textbf{Hybrid 1}:  Same as hybrid 0 except $q \leftarrow \text{LossySamp}(1^{\lambda})$, and $H_{\lambda}= h\circ f_q$. 
    \begin{claim}
    For any QPT adversary $\cal{A}$, there exists a negligible function $\lambda$ such that
    \[
    \Big|\Pr[\mathcal{A}(H_0)=1]-\Pr[\mathcal{A}(H_1)=1]\Big|\leq \text{negl}(\lambda) 
    \]
    
\end{claim}
\begin{proof}
     This follows directly because  for any QPT adversary $\cal{A_{\lambda}}$, 
        \[
        \Big|\Pr_{q \leftarrow \text{InjSamp}(1^{\lambda})}[\mathcal{A_{\lambda}}(q)\rightarrow 1]- \Pr_{q \leftarrow \text{LossySamp}(1^{\lambda})}[\mathcal{A_{\lambda}}(q)\rightarrow 1]\Big|\leq \negl(\lambda)
        \]
\end{proof}

\noindent \textbf{Hybrid 2}: Same as hybrid 1 except the state $\frac{1}{2^{kn/2}}\sum_{x_1, \dots x_n}\ket{x_1}\ket{H(x_1)}\dots \ket{x_n}\ket{H(x_n)}$ is measured on \\$(\sum_{k \in S_1}f(x_k), \dots,\sum_{k \in S_j}f(x_k))$.\\
\begin{claim}
    For any QPT adversary $A$, there exists a negligible function $\lambda$ such that
    \[
    \Big|\Pr[A(H_2)=1]-\Pr[A(H_1)=1]\Big|\leq \text{negl}(\lambda) 
    \]
    
\end{claim}

\begin{proof}
First, some notation: we denote 
\[
\left (\sum_{i\in S_1}f_q(x_i),\sum_{i\in S_2}f_q(x_i),\dots \sum_{i\in S_j}x_i \right)
\]
by $x$, 
\[
\left(\sum_{i\in S_1}f_q(x_i'),\sum_{i\in S_2}f_q(x_i'),\dots \sum_{i\in S_j}f_q(x_i')\right)
\]
by $x'$, \[\left(\sum_{i\in S_1}h(f(x_i)),\sum_{i\in S_2}h(f(x_i)),\dots \sum_{i\in S_j}h(f(x_i))\right)
\]
by $y$, and 
\[
\left(\sum_{i\in S_1}h(f(x_i')),\sum_{i\in S_2}h(f(x_i')),\dots \sum_{i\in S_j}h(f(x_i'))\right)
\]
by $y'$. We let $H_{\lambda}$ be $H$.
Following the proof of Lemma \ref{lem:MH-inj}, we have that,  $\forall x\neq x': \Pr_{h}[\matM^{T}(y-y')=0]= \frac{1}{N}$.
So from the union bound, 
\begin{align*}
    &\Pr_H[\exists x\neq x' \in \text{dist}(\{0,1\}^{k-r},n) : M^T\cdot (y-y')= 0 ]\\
    &\leq 2^{2n(k-r)}\Pr_H[ M^T\cdot( y-y')= 0 ]\\
&=2^{2n(k-r)}\frac{1}{N}\\
&\leq \frac{2^{\log (N)/2}}{N}\\
&\leq \frac{1}{\sqrt{N}}
\end{align*}

\begin{align*}
    &\Pr[E] \leq \Pr[E|x,x'\in \text{dist}(\{0,1\}^{k-r},n\})]\cdot \Pr[\text{dist}(\{0,1\}^{k-r},n\})]+\\ &\quad \Pr[E|x,x'\notin\text{dist}(\{0,1\}^{k-r},n\}) ]\cdot \Pr[x,x'\notin\text{dist}(\{0,1\}^{k-r},n\})]\\
    &\leq \Pr[E|x,x'\in\text{dist}(\{0,1\}^{k-r},n\})]+ \Pr[x,x'\notin\text{dist}(\{0,1\}^{k-r},n\})]\\
    &\leq   \frac{1}{\sqrt{N}} + (\max_{x_i}{W_{x_i}})^n\\
\end{align*}
where $W_{x_i}$ denotes probability of $x_i$ being sampled from the domain distribution.
Now, observe that,  with overwhelming probability over the choice of $q$, $\forall s \in Im(f_q)$,  $\Pr_x[f_q(x)=s]$ must be $\negl(\lambda)$, because otherwise there is a trivial distinguisher between $q \leftarrow \text{LossySamp}(1^{\lambda})$ and  $q \leftarrow \text{InjSamp}(1^{\lambda})$ (Given $q \leftarrow \text{LossySamp}\cup\text{InjSamp}$, the distinguisher can simply evaluate $f_q$ on randomly chosen points, output lossy if it sees repeated outputs, and injective otherwise). This implies that, the weight of every input point $h$ is evaluated on is still $\negl(\lambda)$.

\end{proof}

\noindent \textbf{Hybrid 3}: Same as hybrid 2 except the state $\sum_{x_1, \dots x_n}\ket{x_1}\ket{H(x_1)}\dots \ket{x_n}\ket{H(x_n)}$ is measured on $(\sum_{g \in S_1}h(f(x_g)), \dots, \sum_{g \in S_j}h(f(x_g)))$.\\
\begin{claim}
    For any QPT adversary $A$, there exists a negligible function $\lambda$ such that, with overwhelming probability over $H_{\lambda}$,
    \[
    \Big|\Pr[A(H_2)=1]-\Pr[A(H_3)=1]\Big|\leq \text{negl}(\lambda) 
    \]
    
\end{claim}
\begin{proof}
Let $z=z_1, \dots z_n$, $z'=z_1', \dots z_n', \rho=\sum_{g \in S_1}h(z_g), \dots,\sum_{g \in S_j}h(z_g), \rho'=\sum_{g \in S_1}h(z'_g), \dots,\sum_{g \in S_j}h(z'_g)$.\\
We will prove that, for $i \in [j]$, $h(f(x_1))+\dots h(f(x_{|S_i|}))$ is injective on $(f(x_1), \dots ,f(x_{|S_i|}))$ upto reordering. For ease of notation, let $|S_i|=i$ and let $z_i=f(x_i)$.

 \begin{align*}
     &\Pr[ \exists z_1', \dots z_i' \neq z_1, \dots z_i \in Im(f): h(z_1')+\dots +h(z_i')=h(z_1)+\dots +h(z_i)]\\
     &\leq 2^{2(k-r) i}\cdot  \Pr[  \sum_{g=1}^{i}h(z_g')-\sum_{g=1}^{i}h(z_g)=0]\\
     &= \frac{2^{2(k-r) i}}{N}\\ 
     &\leq \frac{2^{2n(k-r)}}{N}\\
     &\leq \negl(\lambda)
 \end{align*}
where the last inequality follows because $2n(k-r)\leq \frac{1}{4}\log{N}$.
 From the union bound,
 \begin{align}
   &\Pr[\exists (z\neq z':\rho=\rho') \\&\leq \frac{j2^{2n(k-r)} }{N}\\&\leq \frac{n 2^{2n(k-r)}}{N}\\
   &\leq \negl(\lambda)
 \end{align}
\end{proof}

\noindent \textbf{Hybrid 4}: Same as hybrid 3 except  except $q \leftarrow \text{InjSamp}(1^{\lambda})$, and $H_{\lambda}= h\circ f_q$.\\
\begin{claim}
    For any QPT adversary $A$, there exists a negligible function $\lambda$ such that
    \[
    \Big|\Pr[A(H_3)=1]-\Pr[A(H_4)=1]\Big|\leq \text{negl}(\lambda) 
    \]
    
\end{claim}
\begin{proof}
   This follows directly because  for any QPT adversary $\cal{A_{\lambda}}$, 
        \[
        \Big|\Pr_{q \leftarrow \text{InjSamp}(1^{\lambda})}[\mathcal{A_{\lambda}}(q)\rightarrow 1]- \Pr_{q \leftarrow \text{LossySamp}(1^{\lambda})}[\mathcal{A_{\lambda}}(q)\rightarrow 1]\Big|\leq \negl(\lambda)
        \]
\end{proof}
\end{proof}
Thus, we have proven that, for any QPT adversary $\cal{A}$, $\textbf{G}$ is secure. Theorem \ref{thm:lossy} follows directly from Lemma \ref{lem:densmatrix}.
\end{proof}

%% file: RO.tex
\subsection{GMP in the Random Oracle Model}
\label{RO}
 In this section, we show that the GMP assumption holds for a compressing random oracle, unconditionally with query-bounded security. This construction is also orthogonal.
First, some notation: Given 2 sets $\cal{X}$ and $\cal{Y}$, define $\cal{Y}^{\cal{X}}$ as the set of functions $f: \cal{X}\rightarrow \cal{Y}$. Given a distribution $\cal{D}$ on  the set $\cal{Y}$, and another set $\cal{X}$, define $\cal{D}^{\cal{X}}$ as the distribution on $\cal{Y}^{\cal{X}}$ where the output for each distribution is chosen according to $\cal{D}$. The distance between 2 distributions $\mathcal{D}_1$ and $\mathcal{D}_2$ over a set $\cal{T}$ is 
\[
\Big|\mathcal{D}_1-\mathcal{D}_2\Big|= \sum_{t\in \cal{T}}\Big|\mathcal{D}_1(t)-\mathcal{D}_2(t)\Big|
\]
If $\Big|\mathcal{D}_1-\mathcal{D}_2\Big|\leq \epsilon$, then we say $\mathcal{D}_1$ and $\mathcal{D}_2$ are $\epsilon$ close. \\
Let $\textbf{SR}_r^{\calD}= g\circ f$, where $f$ is a random function from $\cal{X}$ to $[r]$, and $g$ is another random function from $[r]$ to $\cal{Y}$ such that $r << \cal{Y}$. \\ 
Now, we will recall the following result from \cite{FOCS:Zhandry12}:
\begin{lemma}
\label{lem:smallrangedist}
    The output distributions of a quantum algorithm making $q$ quantum queries to an oracle either drawn from $\textbf{SR}_r^{\calD}$ or $\cal{D}^{\cal{X}}$ are $\ell(q)/r$-close in total variation distance, where $\ell(q)= \frac{\pi^2(2q)^3}{3}<27q^3$. 
\end{lemma}
With this lemma in hand, we are ready to prove that the generalized matrix problem is hard in the random oracle model. 
\begin{lemma}
  The quantum group action in Construction \ref{constr:hashedbased} instantiated with the random oracle $H_{\lambda}:\{0,1\}^k\rightarrow \Z_{N(\lambda)}$ yields an orthogonal group action.
\end{lemma}
\begin{proof}

 Since $\Pr[H(x)=y]=\frac{1}{N}$  when $H$ is modelled as a random oracle, we can conclude that the group action instantiated with $H$ is orthogonal directly from Lemma \ref{lem:orthogonality}.
   
   \qed \end{proof}

 \begin{theorem}\label{thm:ro}Fix a security parameter $\lambda$, and let $H_{\lambda}:\{0,1\}^k \rightarrow \Z_{N(\lambda)}$ \footnote{From now on, we will use $N$ for $N(\lambda)$, and $H$ for $H(\lambda)$} be the random oracle . Let $m,n$ be polynomials in $\lambda$, and let $N$ be a prime such that $n \leq \frac{\sqrt{\log N}}{4}$. Let $\calM_\lambda$ be a distribution over  $\Z_{N}^{n\times m}$ and let $\calS_\lambda$ be a distribution over $\Z_{N}^{m}$. Then, for all QPT adversaries $\cal{A}$ the $(\calM,\calS)$-Generalized Matrix (GM) Problem is hard in a quantum group action in the random oracle model. 
\end{theorem}

\begin{proof}
From Lemma \ref{lem:densmatrix}, it suffices to prove that the security game $\textbf{G}$ is secure. Formally, we will prove the following lemma:
\begin{lemma}

  Fix a security parameter $\lambda$. For any QPT Adversary $\cal{A}$  with oracle access to the random oracle $H$ in $\textbf{G}$, there exists a negligible function $\text{negl}$ such that
    \[
    \Pr[1\leftarrow Expt^{\cal{A}}(1^{\lambda},1)]-\Pr[1\leftarrow Expt^{\cal{A}}(1^{\lambda},0)]\leq \text{negl} (\lambda)
    \]
\end{lemma}
\begin{proof}
 We will introduce the following sequence of hybrids:\\
\textbf{Hybrid 0}: This is the same as $Expt^{\cal{A}}(1^{\lambda},1)$.\\
\textbf{Hybrid 1}: This is the same as \textbf{Hybrid 0} except the function $H_{\lambda}:\{0,1\}^k\rightarrow \Z_{N}$ is replaced with $f\circ g$, where $g$ is a random function from $\{0,1\}^k$ to $[r]$, $f$ is a random function from $[r]$ to $\Z_{N}$, and $r^{2n}\leq \frac{\sqrt{N}}{4}$. 
\begin{claim}
    For any QPT adversary $\cal{A}$ making $q$ queries to the oracle, there exists a negligible function $\lambda$ such that, 
    \[
    \Big|\Pr[A(H_0)=1]-\Pr[A(H_1)=1]\Big|\leq \frac{\pi^2(2q)^3}{3r}=\frac{4\pi^2(2q)^3}{3N^{1/4n}}=\frac{4\pi^2(2q)^3}{3\cdot 2^{4n}}
    \]
\end{claim}
\begin{proof}
    This directly follows from Lemma \ref{lem:smallrangedist}.
\qed \end{proof}
  \textbf{Hybrid 2}: This is the same as \textbf{Hybrid 1} except the state  is measured on \\$(\sum_{k \in S_1}f(x_k), \dots,\sum_{k \in S_j}f(x_k))$.\\
\begin{claim}
    For any QPT adversary $A$, there exists a negligible function $\lambda$ such that
    \[
    \Big|\Pr[A(H_1)=1]-\Pr[A(H_2)=1]\Big|\leq \text{negl}(\lambda) 
    \]
    
\end{claim}
\begin{proof}
   First, some notation: we denote 
\[
\left (\sum_{i\in S_1}f(x_i),\sum_{i\in S_2}f(x_i),\dots \sum_{i\in S_j}f(x_i) \right)
\]
by $x$, 
\[
\left(\sum_{i\in S_1}f(x_i'),\sum_{i\in S_2}f(x_i'),\dots \sum_{i\in S_j}f(x_i')\right)
\]
by $x'$, \[\left(\sum_{i\in S_1}g(f(x_i)),\sum_{i\in S_2}g(f(x_i)),\dots \sum_{i\in S_j}g(f(x_i))\right)
\]
by $y$, and 
\[
\left(\sum_{i\in S_1}g(f(x_i')),\sum_{i\in S_2}g(f(x_i')),\dots \sum_{i\in S_j}g(f(x_i'))\right)
\]
by $y'$.
Following the proof of Lemma \ref{lem:MH-inj}, we have that,  $\forall x\neq x': \Pr_{h}[\matM^{T}(y-y')=0]= \frac{1}{N^m}$.
So from the union bound, 
\begin{align*}
    &\Pr_H[\exists x\neq x' \in \text{dist}([r],n) : M^T\cdot (y-y')= 0 ]\\
    &\leq r^{2n}\Pr_H[ M^T\cdot( y-y')= 0 ]\\
&=r^{2n}\frac{1}{N}\\
&\leq \frac{1}{4\sqrt{N}}
\end{align*}

\begin{align*}
    &\Pr[E] \leq \Pr[E|x,x'\in \text{dist}([r],n\})]\cdot \Pr[\text{dist}([r],n\})]+\\ &\quad \Pr[E|x,x'\notin\text{dist}([r],n\}) ]\cdot \Pr[x,x'\notin\text{dist}([r],n\})]\\
    &\leq \Pr[E|x,x'\in\text{dist}([r],n\})]+ \Pr[x,x'\notin\text{dist}([r],n\})]\\
    &\leq  \frac{1}{4\sqrt{N}} + (\max_{x_i}{W_{x_i}})^n\\
\end{align*}
where $W_{x_i}$ denotes probability of $x_i$ being sampled from the domain distribution.
Since $W_{x_i}=\frac{1}{r}= \frac{4}{N^{1/4n}}\leq  \frac{4}{2^{4n}}$ for all $x_i \in [r]$, the final quantity is $\negl(\lambda)$.
\qed\end{proof}
\textbf{Hybrid 3}: This is the same as \textbf{Hybrid 2} except the state   is measured on \\$(\sum_{k \in S_1}H(x_k), \dots,\sum_{k \in S_j}H(x_k))$.\\
 \begin{claim}
    For any  adversary $A$, there exists a negligible function $\lambda$ such that
    \[
    \Big|\Pr[A(H_2)=1]-\Pr[A(H_3)=1]\Big|\leq \text{negl}(\lambda) 
    \]
    
\end{claim}

\begin{proof}
Let $x=x_1, \dots x_n$, $x'=x_1', \dots x_n'$, \[\rho=\Big(\sum_{g \in S_1}H(x_g), \dots,\sum_{g \in S_j}H(x_g)\Big),\]  \[\rho'=\Big(\sum_{g \in S_1}H(x'_g), \dots,\sum_{g \in S_j}H(x'_g)\Big),\] \\
We will prove that with high probability over the choice of $H$, $\forall i \in [j]$, $\rho_i$ is injective on $(x_1, \dots ,x_{|S_i|})$ upto reordering. For ease of notation, let $|S_i|=i$.

 \begin{align*}
     &\Pr_{H}[ \exists x_1', \dots x_i' \neq x_1, \dots x_i : \rho_i=\rho_i']\\
     &= \Pr_{f,g}[ \exists x_1', \dots x_i' \neq x_1, \dots x_i: \sum_{u=1}^{i}f(g(x_i'))=\sum_{u=1}^{i}f(g(x_i))]\\
     &= \Pr_{f}[ \exists z_1', \dots z_i' \neq z_1, \dots z_i: \sum_{u=1}^{i}f(z_i)=\sum_{u=1}^{i}f(z_i')]\\
     &\leq r^i\cdot  \Pr[  \sum_{k=1}^{i}f(z_k')-\sum_{k=1}^{i}f(z_k)=0]\\
     &\leq  r^n\cdot  \Pr[  \sum_{k=1}^{i}f(z_k')-\sum_{k=1}^{i}f(z_k)=0]\\
     &= \frac{r^n}{N} 
 \end{align*}
 From the union bound,
 \begin{align}
   \Pr[\exists x\neq x':\rho=\rho'] \\&\leq \frac{j r^{2n}}{N}
   &\leq \frac{n r^{2n}}{N} 
   &=\negl(\lambda)
 \end{align}
 where the last equality follows because $r^{2n}=\frac{\sqrt{N}}{4}$. 
\end{proof}
\textbf{Hybrid 4}: This is the same as \textbf{Hybrid 3} except the function $f\circ g$ is replaced by $H:\cal{X}\rightarrow \cal{Y}$, which is the random oracle. 
\begin{claim}
    For any QPT adversary $\cal{A}$ making $q$ queries to the oracle, there exists a negligible function $\lambda$ such that, 
    \[
    \Big|\Pr[A(H_4)=1]-\Pr[A(H_3)=1]\Big|\leq \frac{\pi^2(2q)^3}{3r}=\frac{4\pi^2(2q)^3}{3N^{1/4n}}=\frac{4\pi^2(2q)^3}{3\cdot 2^{4n}}
    \]
\end{claim}
\begin{proof}
    This directly follows from Lemma \ref{lem:smallrangedist}.
\end{proof}

\end{proof}
\end{proof}

%% file: structuredGMP.tex
\section{Structured GMP}\label{sec:gmp2}

In this section, we will present and prove security of a more structured version of the Generalized Matrix Problem. 
\begin{theorem}
\label{thm:strucgmpexpanding}
Fix a security parameter $\lambda$. Let $k, n$ be polynomials in $\lambda$, and let $N$ be a prime such that $n \leq (\log N)^{1/4}$.
Let $(\mathbb{G},\cal{X},*)$ be a hash based quantum-state group action instantiated with the $H_{\lambda}$.  Let $(\cal{M}_{\lambda} )$ be a distribution over matrices $\matM \in \Z_{N}^{n\times m}$ and let $(\cal{S}_{\lambda} )$ be a distribution over vectors $\vecs\in  \{0,1\}^{m}$ such that for all $\epsilon\in [0,1]$ at least $1-\sqrt{\epsilon}$ fraction of matrices $\matM$ satisfy the following:\\
For all $x_1, \dots x_n, x_1', \dots x_n'$ such that $(H_{\lambda}(x_1)-H_{\lambda}(x_1'), \dots H_{\lambda}(x_n)-H_{\lambda}(x_n')) \neq 0$,
\[\frac{1}{2}(\sum_{z\in \Z_N}\Big|\Pr_{\vecs}[(H_{\lambda}(x_1)-H_{\lambda}(x_1'), \dots H_{\lambda}(x_n)-H_{\lambda}(x_n'))\cdot \matM\cdot \vecs=z]-\frac{1}{N}\Big|)\leq \sqrt{\epsilon}
\] 
Then, with probability at least $1-\sqrt{\epsilon}-\frac{1}{2^{kn}}$ over the choice of $M,H$, the Generalized Matrix Problem is hard in the hash based quantum group action instantiated with the the function $H: \{0,1\}^{k} \rightarrow \Z_N$ which is either the random oracle, or is a function from Construction \ref{expandingcons} or \ref{lossycons}.
 
\end{theorem}
\begin{proof}

    When $b=1$, $\cal{A}$ receives $\{\ket{\psi}_{v^{(1)}_1},\ket{\psi}_{v^{(1)}_2}, \dots , \ket{\psi}_{v^{(1)}_n}\} $, where $v^{(1)}_i=(\matM\cdot\vecs)_i$. The density matrix representing the corresponding mixed state is:  
\begin{align*}
&\sum_{\vecs}\Pr[\vecs] (\matM\cdot \vecs)_1*\ket{\psi}_0\bra{\psi}_0*(\matM\cdot \vecs)_1\otimes \dots \otimes (\matM\cdot \vecs)_n*\ket{\psi}_0\bra{\psi}_0*(\matM\cdot \vecs)_n\\
&= \frac{1}{2^{kn}}\sum_{\vecs,x_1,x_1', \dots x_n, x_n'}\Pr[\vecs]\omega_N^{H(x_1)-H(x_1')\cdot (\matM\cdot \vecs)_1}\ket{x_1}\bra{x_1'}\otimes \dots \otimes \omega_N^{H(x_n)-H(x_n')\cdot (\matM\cdot \vecs)_n }\ket{x_n}\bra{x_n'}\\
&= \frac{1}{2^{kn}}\sum_{\vecs,x_1,x_1', \dots x_n, x_n'}\Pr[\vecs]\omega_N^{(H(x_1)-H(x_1'), \dots H(x_n)-H(x_n'))\cdot (\matM\cdot \vecs)}\ket{x_1}\bra{x_1'}\otimes \dots \otimes \ket{x_n}\bra{x_n'}
\end{align*}
 For simplicity of notation, let $x=x_1,\dots, x_n$, $x'= x_1', \dots,x_n'$ ,$\mu_{x}= (H(x_1), \dots H(x_n))\cdot \matM, \mu_{x'}=(H(x_1'), \dots H(x_n'))\cdot \matM,\mu_{x-x'}=(H(x_1')-H(x_1), \dots H(x_n)-H(x_n'))\cdot \matM$.
    Now recall that we have already shown that,   
    \begin{align*}
      &\bigg|\frac{1}{2^{kn}}\Pr[\mathcal{A}^H(\sum_{x,x':\mu_{x}=\mu_{x'}}\ket{x}\bra{x'})=1]\quad \quad\quad\\ &\;\;\;-
    \frac{1}{2^{kn}}\Pr[\mathcal{A}^H(\sum_{\substack{x,x':\\(H(x_1), \dots H(x_n))=\\(H(x'_1), \dots H(x'_n))}}\ket{x}\bra{x'})=1]\bigg|\leq \text{negl}(\lambda)  
    \end{align*}
    when $H$ is the random oracle or a function from Construction \ref{expandingcons}. In the case when $H$ is a function from Construction \ref{lossycons}, we have shown that $\frac{1}{2^{kn}}\sum_{x,x':\mu_{x}=\mu_{x'}}\ket{x}\bra{x'}$ and $\frac{1}{2^{kn}}\sum_{\substack{x,x':\\(H(x_1), \dots H(x_n))=\\(H(x'_1), \dots H(x'_n))}}\ket{x}\bra{x'}$ are computationally indistinguishable.

  So, if we show that $\rho'=\frac{1}{2^{kn}}\sum_{x,x':\mu_{x}=\mu_{x'}}\ket{x}\bra{x}'$ and\\ $\rho=\frac{1}{2^{kn}}\sum_{\vecs,x,x'}\Pr[\vecs]\omega_N^{(H(x_1)-H(x_1'), \dots, H(x_n)-H(x_n'))\cdot (M\cdot \vecs)}\ket{x}\bra{x'}$
  are close in trace distance, we would be done.
 \begin{align*}
      \rho &= \frac{1}{2^{kn}}\sum_{\vecs,x,x'}\Pr[\vecs]\omega_N^{\langle \mu_{x-x'},  \vecs\rangle}\ket{x}\bra{x'}\\
      &= \frac{1}{2^{kn}}\sum_{\vecs}\Pr[\vecs]\Big(\sum_{x,x':\mu_{x}=\mu_{x'}}\omega_N^{\langle \mu_{x-x'},  \vecs\rangle}\ket{x}\bra{x'}+\sum_{x,x':\mu_{x}\neq \mu_{x'}}\omega_N^{\langle \mu_{x-x'},  \vecs\rangle}\ket{x}\bra{x'}\Big)\\
      &= \frac{1}{2^{kn}}\sum_{\vecs}\Pr[\vecs]\Big(\sum_{x,x':\mu_{x}=\mu_{x'}}\ket{x}\bra{x'}+\sum_{x,x':\mu_{x}\neq \mu_{x'}}\omega_N^{\langle \mu_{x-x'},  \vecs\rangle}\ket{x}\bra{x'}\Big)\\
      &= \frac{1}{2^{kn}}\sum_{x,x':\mu_{x}=\mu_{x'}}\ket{x}\bra{x'}+\frac{1}{2^{kn}}\sum_{x,x':\mu_{x}\neq \mu_{x'}}\ket{x}\bra{x'}\sum_{\vecs}\Pr[\vecs]\omega_N^{\langle \mu_{x-x'},  \vecs\rangle} \\ 
      &=\rho'+ \frac{1}{2^{kn}}\sum_{x,x':\mu_{x}\neq \mu_{x'}}\ket{x}\bra{x'}\sum_{\vecs}\Pr[\vecs]\omega_N^{\langle \mu_{x-x'},  \vecs\rangle}
  \end{align*}

Now, let $\matM$ be $\sqrt{\epsilon}$ "good" if, for all $x_1, \dots x_n, x_1', \dots x_n'$ such that $(H(x_1)-H(x_1'), \dots H(x_n)-H(x_n')) \neq 0$\[\frac{1}{2}(\sum_{z\in \Z_N}\Big|\Pr_{\vecs}[(H(x_1)-H(x_1'), \dots H(x_n)-H(x_n'))\cdot \matM\cdot \vecs=z]-\frac{1}{N}\Big|)\leq \sqrt{\epsilon},
\]  and let $(\rho-\rho')_{x,x'}=\frac{1}{2^{kn}}\sum_{\vecs}\Pr[\vecs]\omega_N^{(H(x_1)-H(x_1'), \dots H(x_n)-H(x_n'))\cdot (\matM\cdot\vecs)}$. 
\begin{claim}
   For all $\sqrt{\epsilon}$-good $\matM$, for all $x,x'$ such that $(H(x_1)-H(x_1'), \dots H(x_n)-H(x_n'))\neq 0: |(\rho-\rho')_{x,x'}|\leq \frac{2\sqrt{\epsilon}}{2^{nk}}$
\end{claim}
\begin{proof}
    Fix a $\sqrt{\epsilon}$-good $\matM$, and let $x, x'$ be such that $(H(x_1)-H(x_1'), \dots, H(x_n)-H(x_n'))\neq 0$. Define $p_z=\Pr_{\vecs}[\langle \mu_{x-x'},\vecs\rangle =z]$ and $\epsilon_z=p_z-\Pr[U_{N}=z]=p_z-1/N$. Recall that  $ \frac{1}{2}\sum_{z\in [N]}  |\epsilon_z|\leq \sqrt{\epsilon}$. \\
   Then \begin{align*}
\left|\sum_{\vecs}\Pr[\vecs]\omega_N^{\langle \vecs,\mu_{x-x^*}\rangle} \right|
    &= \left|\sum_z p_z\omega_N^z\right|
    = \left|\sum_z \frac{1}{N}\omega_N^z + \sum_z\epsilon_z\omega_N^z\right|
    = \left|\sum_z\epsilon_z\omega_N^z\right|\\
    &\leq \sum_z |\epsilon_z\omega_N^z|
    =\sum_z|\epsilon_z|\leq2\sqrt{\epsilon}
\end{align*}

Now recall that \[\rho-\rho'=\frac{1}{2^{kn}}\sum_{x,x':\mu_{x}\neq \mu_{x'}}\ket{x}\bra{x'}\sum_{\vecs}\Pr[\vecs]\omega_N^{\langle \mu_{x-x^*},  \vecs\rangle}.\] Thus  $|(\rho-\rho')_{x,x'}|\leq  \frac{2\sqrt{\epsilon}}{2^{kn}} $. \qed\end{proof}

Now we have that \[||\rho-\rho'||_{td}= \sum_{\lambda: \text{ eigenvalues of } \rho-\rho'}|\lambda|\]
Let $\sigma$ be the vector of the singular values of $\rho-\rho'$, let $M$ be a $\sqrt{\epsilon}$ good matrix, and let $G$ denote the set of all $x,x'$ such that \\$(H(x_1)-H(x_1'), \dots ,H(x_n)-H(x_n'))\neq 0$.
\begin{align*}
    ||\rho-\rho'||_{td}&= ||\sigma||_1
    \leq 2^{nk/2}||\sigma||_2
    \leq 2^{nk/2} || \rho-\rho'||_F\\
    &\leq 2^{nk/2}\cdot \sqrt{\sum_{x \neq x'}|(\rho-\rho')_{x,x'}^2|}\\
&=2^{nk/2}\cdot\Big(\sqrt{\sum_{\ x \neq x'\in G}|(\rho-\rho')_{x,x'}|^2+\sum_{ x \neq x'\notin G}|(\rho-\rho')_{x,x'}|^2}\Big) \\
&\leq 2^{nk/2}\cdot \sqrt{|G|\frac{4\epsilon}{2^{2kn}}+(2^{2kn}-|G|)\frac{1}{2^{2kn}}}\\
&= \frac{1}{2^{nk/2}}\sqrt{|G|4\epsilon+(2^{2kn}-|G|)}\\
&= \frac{1}{2^{nk/2}}\sqrt{|G|(4\epsilon-1)+2^{2kn}}
\end{align*}
Now, since $H$ is $2n$ wise independent, 
\begin{align*}
    \Pr[(H(x_1)-H(x_1'), \dots ,H(x_n)-H(x_n'))=\vec{0}]&= \prod_{i=1}^{n}\Pr[H(x_i)-H(x_i')=0]\\
    &= \frac{1}{N^n}
\end{align*}

By Markov's inequality,
\begin{align*}
    &\Pr[|x,x':(H(x_1)-H(x_1'), \dots H(x_n)-H(x_n'))=\vec{0}|\geq 2^{2kn}\cdot \epsilon']\\
    &\leq \frac{2^{2kn}}{N^n}\cdot\frac{1}{2^{2kn}\cdot \epsilon'}
    \leq\frac{1}{2^{4kn}}\cdot \frac{1}{\epsilon'}\\
    \end{align*}
   Let $\epsilon'=\frac{1}{2^{2kn}}$. Then, with probability at least $1-\frac{1}{2^{6kn}}$,
    \begin{align*}
       ||\rho-\rho'||_{td} 
       &= \frac{1}{2^{nk/2}}\sqrt{|G|\cdot (4\epsilon)+2^{2kn}-|G|}\\
       &= \frac{1}{2^{nk/2}}\sqrt{2^{2kn}\cdot 4\epsilon}
       = \frac{2^{nk/2}\sqrt{\epsilon}}{2}
    \end{align*}
  \begin{remark}
      Observe that, in the case when $H$ is a function from construction \ref{lossycons}, since $f_q$ is injective, the density matrix $\frac{1}{2^k}\sum_{x}\omega^{H(x)\cdot g}\ket{x}\bra{x}$, where $g \in \Z_N$, can be rewritten as $\frac{1}{2^k}\sum_{x}\omega^{h( f_q(x))\cdot g}\ket{x}\bra{x}$. Switching to $q \leftarrow Lossysamp$, this density matrix looks indistinguishable from  $\frac{1}{2^k}\sum_{x}\omega^{h(u)\cdot g}\ket{x}\bra{x}$, where $u =f_q(x)$. We can rewrite this matrix as
      \[
      \frac{1}{[Im(f_q)]}\sum_{u \in [Im(f_q)]}\omega^{h(u)\cdot g}\sum_{x:f_q(x)=u}\ket{x}\bra{x}
      \]
      Doing the same analysis on this density matrix, we can get a better bound on the trace distance of $\rho-\rho'$, because the dimension of the density matrix now depends on $|Im(f_q)|$, where $q\leftarrow Lossysamp$.\\
      A similar analysis would work when switching from the random oracle to $\textbf{SR}_r^{\calD}= g\circ f$.

  \end{remark}     
\end{proof}

%% file: LHS.tex
\section{LHS Assumption}\label{lhs}
In this section, we show that the LHS assumption is a special case of the structured GMP assumption, and also prove that it is hard in the quantum group action model, for a limited number of samples.

\begin{theorem}
Let $k, n$ be polynomials in $\lambda$, and let $N$ be a prime such that $n \leq (\log N)^{1/4}$, and let $\epsilon\in (0,1]$.
Let $(\mathbb{G},\cal{X},*)$ be a hash based quantum-state group action.
  Let $\calM=(\calM_\lambda)_\lambda$ be a uniform distribution over matrices $\matM\in \Z_N^{n \times m}$ ,  and let $\calS=(\calS_\lambda)_\lambda$ be a distribution over $\{0,1\}^{m}$ such that $H_{\infty}(\calS)=\log N-2\log\epsilon$. 
Then, for all QPT adversaries $\cal{A}^{H}$, with probability at least $1-\sqrt{\epsilon}-\frac{1}{2^{kn}}$ over the choice of $M,H$, the Generalized Matrix Problem is hard in the hash based quantum group action instantiated with the the function $H: \{0,1\}^{k} \rightarrow \Z_N$ which is either the random oracle, or is a function from Construction \ref{expandingcons} or \ref{lossycons}.
\end{theorem}

   \begin{proof}
   From theorem \ref{thm:strucgmpexpanding}, it suffices to prove the following:
   \begin{lemma}
   At least $1-\sqrt{\epsilon}$ fraction of matrices $\matM \in {\sf EqPat}_\matM(\Z_N^{n\times m})$ satisfy the following:\\
For all $x_1, \dots x_n, x_1', \dots x_n'$ such that $(H(x_1)-H(x_1'), \dots H(x_n)-H(x_n')) \neq 0$,
\[\frac{1}{2}(\sum_{z\in \Z_N}\Big|\Pr_{\vecs}[(H(x_1)-H(x_1'), \dots H(x_n)-H(x_n'))\cdot \matM\cdot \vecs=z]-\frac{1}{N}\Big|)\leq \sqrt{\epsilon}
\] 
 \end{lemma}
\begin{proof}
   
    Recall that, for each of the constructions (Construction \ref{expandingcons}, Construction \ref{lossycons}, R.O) of $H$, for fixed $x_1, \dots x_n, x_1', \dots x_n'$:
    \begin{align*}
        \Pr_{H}[(H(x_1)-H(x_1'), H(x_2)-H(x_2'), \dots H(x_n)-H(x_n)'=y_1, \dots y_n]&= \frac{1}{N^n}.
    \end{align*}
   This follows because $H$ is $2n$ wise independent in each of the constructions. 
  Now, for a randomly chosen $\matM$, and for all $x,x'$ such that $(H(x_1)-H(x_1'), \dots ,H(x_n)-H(x_n'))\neq 0$,
        $(H(x_1)-H(x_1'), \dots H(x_n)-H(x_n'))\cdot \matM$ is a uniformly random vector. 
 From now on, we will denote $(H(x_1)-H(x_1'), \dots H(x_n)-H(x_n')) $ by $z_{x,x'}$.
Fix  $
x,x'$ such that $z_{x,x'} \neq 0$, and let  $h^*_{\matM}(s)=\langle z_{x,x'}\cdot M,s\rangle$. Let $H^*$ denote the family of such hash functions. Then from the leftover hash lemma, we can conclude that
\[
d_{\text{TV}}((H^*(\cal{S}),H^*), (U_{N},H^*))\leq \epsilon
\]
where $\epsilon$ is such that $2 \log (\frac{1}{\epsilon})\leq H_{\infty}(\mathcal{S})-\log {N}=q-\log N$. This implies that $\epsilon=\frac{\sqrt{N}}{2^{q/2}}$.
Rewriting the expression from the leftover hash lemma, we have the guarantee that
\begin{align*}
    \mathbb{E}_{H^*\sim H^*}[\Delta(h^*(\cal{S}),U_{N})]&= \mathbb{E}_{\matM \sim \cal{M}}[\Delta(h^*_{\matM}(\calS),U_{N})]\\
    &\leq \frac{\sqrt{N}}{2^{q/2}}
\end{align*} 
\end{proof}

Now, let $\matM$ be $\sqrt{\epsilon}$ "good" if, for all $x_1, \dots x_n, x_1', \dots x_n'$ such that $z_{x,x'} \neq 0$\[\frac{1}{2}(\sum_{u\in \Z_N}\Big|\Pr_{\vecs}[z_{x,x'}\cdot \matM\cdot \vecs=u]-\frac{1}{N}\Big|)\leq \sqrt{\epsilon}.
\]
Let $(\rho-\rho')_{x,x'}=\frac{1}{2^{kn}}\sum_{\vecs}\Pr[\vecs]\omega_N^{z_{x,x'}\cdot (\matM\cdot\vecs)}$. \\
Now, we are ready to prove that
  at least $1-\sqrt{\epsilon}$ fraction of matrices $\matM$ are $\sqrt{\epsilon}$ good. 
  Let $\mathbf{H^*}$ be the random variable distributed uniformly over the support of $\cal{H}^*$. 
  Since $d_{\text{TV}}((H^*(\cal{S}),H^*), (U_{N},H^*))\leq \epsilon$, 
\begin{align*}
    &\frac{1}{2}\sum_{u\in \Z_N, h^*_{\matM}}\Big| \Pr[\mathbf{H^*}(\mathcal{S})=u, \mathbf{H^*}=h^*_{\matM}]-\Pr[U_N=u, U_{|\mathbf{H^*}|}=h^*_{\matM}]\Big|\\
    &= \frac{1}{2}\sum_{u \in \Z_N, h^*_{\matM}}\Big| \Pr[\mathbf{H^*}(\mathcal{S})=u| \mathbf{H^*}=h^*_{\matM}]\Pr[\mathbf{H^*}=h^*_{\matM}]-\Pr[U_{N}=u]\cdot \Pr[ U_{|\mathbf{H^*}|}=h^*_{\matM}]\Big|\\
    &= \frac{1}{2}\sum_{ h^*_{\matM},u}\Pr[\mathbf{H^*}=h^*_{\matM}]\Big| \Pr[\mathbf{H^*}(\mathcal{S})=u| \mathbf{H^*}=h^*_{\matM}]-\Pr[U_{N}=u]\Big|\\
    &=\frac{1}{2}\sum_{ h^*_{\matM},u}\Pr[\mathbf{H^*}=h^*_{\matM}]\Big| \Pr[h^*_{\matM}(\mathcal{S})=u]-\Pr[U_{N}=u]\Big|
 \end{align*} 
 If less than $1-\sqrt{\epsilon}$ fraction of functions $h^*_{\matM}$ are such that $\Big| \Pr[h^*_{\matM}(\mathcal{S})=u]-\Pr[U_{N}=u]\Big|\leq \sqrt{\epsilon}$, then at least $\sqrt{\epsilon}$ fraction of functions $h^*_{\matM}$ are such that $\Big| \Pr[h^*_{\matM}(\mathcal{S})=z]-\Pr[U_{N}=u]\Big|> \sqrt{\epsilon}$, and so $\sum_{ h^*_{\matM},u}\Pr[\mathbf{H^*}=h^*_{\matM}]\Big| \Pr[H^*(\mathcal{S})=u]-\Pr[U_{N}=u]\Big|>\epsilon$, which is a contradiction. \\
 Now, the proof goes through essentially unchanged from Theorem \ref{thm:strucgmpexpanding}.
\end{proof}

%% file: qmoney.tex
\section{Quantum Money Scheme}
\label{qmoney}
Quantum money utilizes the un-clonability of quantum states to protect against couonterfeiting. An important feature of quantum money is \emph{public verifiability}, whereby anyone can verify banknotes, while only the mint can create them. 

Here, we discuss two constructions of quantum money, that of~\cite{EC:Zhandry19b} and that of~\cite{ITCS:Zhandry24a}. At first glance these constructions look completely different. Ultimately, however, we will explain how quantum state group actions allows us to connect the two.

\subsection{Defining Quantum Money and Quantum Lightning} 

This section is taken almost verbatim from~\cite{ITCS:Zhandry24a}. Here we define quantum money and quantum lightning. In the case of quantum money, we focus on \emph{mini-schemes}~\cite{STOC:AarChr12}, which are essentially the setting where there is only ever a single valid banknote produced by the mint. As shown in~\cite{STOC:AarChr12}, such mini-schemes can be upgraded generically to full quantum money schemes using digital signatures. 

\paragraph{Syntax.} Both quantum money mini-schemes and quantum lightning share the same syntax:
\begin{itemize}
	\item $\gen(1^\lambda)$ is a quantum polynomial-time (QPT) algorithm that takes as input the security parameter (written in unary) which samples a classical serial number $\sigma$ and quantum banknote $\$$. 
	\item $\ver(\sigma,\$)$ takes as input the serial number and a supposed banknote, and either accepts or rejects, denoted by $1$ and $0$ respectively.
\end{itemize}

\paragraph{Correctness.} Both quantum money mini-schemes and quantum lightning have the same correctness requirement, namely that valid banknotes produced by $\gen$ are accepted by $\ver$. Concretely, there exists a negligible function $\negl(\lambda)$ such that
\[\Pr[\ver(\sigma,\$)=1:(\sigma,\$)\gets\gen(1^\lambda)]\geq 1-\negl(\lambda)\enspace.\]

\paragraph{Security.} We now discuss the security requirements, which differ between quantum money and quantum lightning.

\begin{definition}Consider a QPT adversary $\calA$, which takes as input a serial number $\sigma$ and banknote $\$$, and outputs two potentially entangled states $\$_1,\$_2$, which it tries to pass off as two banknnotes. $(\gen,\ver)$ is a secure \emph{quantum money mini-scheme} if, for all such $\calA$, there exists a negligible $\negl(\lambda)$ such that the following holds:
\[\Pr\left[\ver(\sigma,\$_1)=\ver(\sigma,\$_2)=1:\substack{(\sigma,\$)\gets\gen(1^\lambda)\\(\$_1,\$_2)\gets\calA(\sigma,\$)}\right]\leq\negl(\lambda)\enspace .\]
\end{definition}
\begin{definition}Consider a QPT adversary $\calB$, which takes as input the security parameter $\lambda$, and outputs a serial number $\sigma$ and two potentially entangled states $\$_1,\$_2$, which it tries to pass off as two banknnotes. $(\gen,\ver)$ is a secure \emph{quantum lightning} scheme if, for all such $\calB$, there exists a negligible $\negl(\lambda)$ such that the following holds:
	\[\Pr\left[\ver(\sigma,\$_1)=\ver(\sigma,\$_2)=1:(\sigma,\$_1,\$_2)\gets\calB(1^\lambda)\right]\leq\negl(\lambda)\enspace .\]
\end{definition}
Quantum lightning trivially implies quantum money: any quantum money adversary $\calA$ can be converted into a quantum lightning adversary $\calB$ by having $\calB$ run both $\gen$ and $\calA$. But quantum lightning is potentially stronger, as it means that even if the serial number is chosen adversarially, it remains hard to devise two valid banknotes. This in particular means there is some security against the mint, which yields a number of additional applications, as discussed by~\cite{EC:Zhandry19b}.

\subsection{Quantum Money from Non-collapsing Hashes}

Here, we briefly recall a result of~\cite{EC:Zhandry19b}, that a certain type of hash function implies quantum money. 

For a quantum state $|\phi\rangle$, let $r\gets|\phi\rangle$ denote the probabilistic process of measuring $|\phi\rangle$ in the computational basis. For a classical function $H$, let $(h,|\tau_h\rangle)\stackrel{H}{\gets}|\phi\rangle$ denote the process of mapping $|\phi\rangle=\sum_r \alpha_r\rangle$ to $\sum_r \alpha_r|r,H(r)\rangle$, measuring $H(r)$ to get $h$, and then outputting the state $|\phi_h\rangle$ that is the state remaining after measurement, conditioned on the measurement outcome being $h$.
\begin{definition}A family $H=(H_\lambda)_\lambda$ of hash functions is non-collapsing if there exists an efficiently constructible family of states $(|\phi_\lambda)_\lambda$ with support on $\calR_\lambda$, and a QPT algorithm $\test$, and negligible function $\negl(\lambda)$ such that:
\begin{itemize}
    \item $\Pr[\test(|r\rangle)=0:r\gets|\phi_\lambda\rangle]\geq 1-\negl(\lambda)$
    \item $\Pr[\test(|\tau_h\rangle)=1:(y,|\tau_h\rangle)\stackrel{H_\lambda}{\gets}|\phi_\lambda\rangle]\geq 1-\negl(\lambda)$
\end{itemize}
\end{definition}
In other words, a non-collapsing hash function allows for distinguishing whether a state is fully measured, or if only the the output of $H$ is measured; in the latter case, the state is a superposition of many pre-images, but looks as though it is just a single classical value.

\begin{construction}[\cite{EC:Zhandry19b}]\label{constr:qmoneyfromhash} Let $H=(H_\lambda)$ be a family of hash functions that are assumed to be non-collapsing. Consider the following quantum money scheme $(\gen,\ver)$:
\begin{itemize}
    \item $\gen(1^\lambda)$: sample $(h,|\tau_h\rangle)\stackrel{H_\lambda}{\gets}|\phi_\lambda\rangle$. Set $\sigma=h$ and $\$=|\tau_h\rangle$.
    \item $\ver(\sigma=h,\$)$: Run $(h',|\tau'_{h'}\rangle)\stackrel{H_\lambda}{\gets}\$$. If $h'\neq h$, immediately abort and output 0. Otherwise, assuming $h'=h$, run $b\gets\test(|\tau'_{h'})$, and output $b$.
\end{itemize}
\end{construction}

In~\cite{EC:Zhandry19b}, it is shown that a collision-resistant non-collapsing $H$ leads to Construction~\ref{constr:qmoneyfromhash} being a secure quantum lightning scheme. More generally, even if the $\test$ algorithm for non-collapsing only has an inverse-polynomial distinguishing probability, then Construction~\ref{constr:qmoneyfromhash} can be adapted int oa secure quantum lightning scheme.

\subsection{Quantum Money from Abelian Group Actions}

Here, we now recall a more recent quantum money scheme due to~\cite{ITCS:Zhandry24a} using group actions. The scheme is initially described using classical group actions.

\begin{construction}[\cite{EC:Zhandry19b}]\label{constr:qmoneyfromaction} Let $(\;(\G_\lambda,\calX_\lambda,*_\lambda)\;)_\lambda$. We will assume for simplicity that $\G_\lambda$ is a cyclic group $\Z_{N(\lambda)}$. We will also drop the subscript $\lambda$ from $*$ to de-clutter the notation. Finally, we will assume the ability to recognize elements in $\calX_\lambda$, distinguishing them from arbitrary strings.
	\begin{itemize}
		\item $\gen(1^\lambda)$: Initialize quantum registers $\calS$ (for serial number) and $\calM$ (for money) to states $|0\rangle_\calS$ and $|0\rangle_\calM$, respectively. Then do the following:
		\begin{itemize}
			\item Apply $\QFT_{N(\lambda)}$ to $\calS$, yielding the joint state $\frac{1}{\sqrt{N(\lambda)}}\sum_{g\in\Z_{N(\lambda)}}|g\rangle_\calS|0\rangle_\calM$.
			\item Apply in superposition the map $|g\rangle_\calS|y\rangle_\calM\mapsto |g\rangle_\calS|y\oplus (g*x_\lambda)\rangle_\calM$. The joint state of the system $\calS\otimes\calM$ is then $\frac{1}{\sqrt{N(\lambda)}}\sum_{g\in\Z_{N(\lambda)}}|g\rangle_\calS|g*x_\lambda\rangle_\calM$.
			\item Apply $\QFT_{N(\lambda)}$ to $\calS$ again, yielding $\frac{1}{N(\lambda)}\sum_{g,h\in\Z_{N(\lambda)}}e^{i2\pi gh/N(\lambda)}|h\rangle_\calS|g*x_\lambda\rangle_\calM$
			\item Measure $\calS$, giving the serial number $\sigma:=h$. The $\calM$ register then collapses to the banknote $\$_h:=\frac{1}{\sqrt{N(\lambda)}}\sum_{g\in\Z_{N(\lambda)}}e^{i2\pi gh/N(\lambda)}|g*x_\lambda\rangle_\calM$. Output $(h,\$_h)$. 
		\end{itemize}
		\item $\ver(\sigma=h,\$):$ First verify that the support of $\$$ is contained in $\calX_\lambda$, by applying the assumed algorithm for recognizing $\calX_\lambda$ in superposition. Then do the following:
		\begin{itemize}
			\item Initialize a new register $\calH$ to $\frac{1}{\sqrt{N(\lambda)}}\sum_{u\in\Z_{N(\lambda)}}|u\rangle_\calH$
			\item Apply in superposition the map $|u\rangle_\calH|y\rangle_\calM\mapsto |u\rangle_\calH|(-u)*y\rangle_\calM$. 
			\item Apply $\QFT^{-1}_{N(\lambda)}$ to $\calH$.
			\item Measure $\calH$, obtaining a group element $h'$. Accept if and only if $h'=h$.
		\end{itemize}
	\end{itemize}
\end{construction}
In~\cite{ITCS:Zhandry24a}, it is shown that verification accepts exactly the state $\$_h$ and all states orthogonal to $\$_h$ are rejected.

\subsection{Generalization to Quantum State Group Actions}

It is relatively straightforward to adapt Construction~\ref{constr:qmoneyfromaction} to quantum state group actions. The result is that the monst state $\$_\sigma$ takes the form
\[\$_h=\frac{1}{\sqrt{N(\lambda)}}\sum_{g\in\Z_{N(\lambda)}} e^{i\pi g h/N(\lambda)}|\psi_{g*x_\lambda}\rangle\]

Now consider implementing the quantum state group action with Construction~\ref{constr:hashedbased}, using a hash function $H_\lambda$. One issue is that Construction~\ref{constr:qmoneyfromaction} requires the ability to recognize set elements, but our Construction~\ref{constr:hashedbased} does not give such an ability. We will return to this issue in a moment.

Recall that the elements $|\psi_{g*x_\lambda}\rangle$ in Construction~\ref{constr:hashedbased} have the form
\[|\psi_{g*x_\lambda}\rangle=\sum_r \alpha_r e^{i2\pi g H_\lambda(r)}|r\rangle\]
Above, $|\phi_\lambda\rangle=\sum_r \alpha_r |r\rangle$ is the starting state of the quantum state group action. Combining with our expression for $\$_h$, we can sum over $g$, giving
\[\$_h=\sqrt{N(\lambda)}\sum_{r:H(r)=h} \alpha_r |r\rangle\]
But these are \emph{exactly} the states $|\tau_h\rangle$ used in the quantum money scheme from collapsing hashes in Construction~\ref{constr:qmoneyfromhash}.

Thus, we see that $\gen$ in Construction~\ref{constr:qmoneyfromhash} is equivalent to $\gen$ in Construction~\ref{constr:qmoneyfromaction} when instantiating the group action using hashes as in Construction~\ref{constr:hashedbased}.

Now, recall that $\ver$ in Construction~\ref{constr:qmoneyfromaction} requires the ability to recognize set elements. Observe that the states $|\tau_h\rangle$ are simply another basis for the span of set elements $|\psi_{g*x_\lambda}\rangle$. Thus, we see that $H$ being non-collapsing corresponds \emph{exactly} to the ability to recognize set elements. Thus, we see that $\ver$ in Construction~\ref{constr:qmoneyfromhash} is also equivalent to $\ver$ in Construction~\ref{constr:qmoneyfromaction}, using this view of $\test$ as recognizing set elements.

Thus, we see that the quantum money construction of~\cite{EC:Zhandry19b} (Construction~\ref{constr:qmoneyfromaction}) is simply an instance of the construction of~\cite{ITCS:Zhandry24a} (Construction~\ref{constr:qmoneyfromaction}), when instantiated using our hash-based quantum state group action from Construction~\ref{constr:hashedbased}.

%% file: qkd.tex
\section{Quantum State Group Actions and QKD}\label{qkd}

The original Diffie-Hellman protocol~\cite{DifHel76} solves the problem of \emph{key distribution}: allowing Alice and Bob to establish a secure key in the presence of an eavesdropper. It is assumed that Alice and Bob have a classical \emph{authenticated} channel at their disposal. It is classically known that such a channel cannot be used to information-theoretically establish a shared key, and computational security and hence unproven computational assumptions are required.

Quantum Key Distribution (QKD)~\cite{BB84} solves the same problem, but uses quantum communication in addition to the classical authenticated channel. Now, it is possible to attain information-theoretic security, which allows for unconditional security proofs.

Here, we give a toy protocol for key distribution from an abelian group action. If the group action is a classical group action on which DDH holds, the resulting scheme is computationally secure. On the other hand, if the group action is an information-theoretic quantum-state group action (such as our group action from Section~\ref{sec:hashbased}), then the resulting scheme is information-theoretically secure QKD. Thus, we can see how (quantum state) group actions can also unify the concepts of classical key distribution and QKD.

Our construction is inspired by the group action version of the Diffie-Hellman protocol. One key challenge is that if the group action is quantum, the shared key is now a quantum state, rather than a classical key. A second challenge is that if the group action is quantum, sending the messages now requires an authenticated \emph{quantum} challenge. But as quantum authentication implies encryption, such a channel makes key distribution unnecessary. So we need a mechanism to incorporate a classical authenticated channel even if the group action is quantum.

We resolve both of these difficulties using standard techniques. The resulting protocol does not appear to offer any advantages over classical or quantum key distribution protocols. But we include it as an interesting conceptual contribution. We briefly give the intuition for security.

\begin{construction}We assume a (possible classical) group action, with underlying states $|\psi_g\rangle$ and group $\G$. The protocol works as follows.
\begin{enumerate}
    \item\label{step:alicefirstmessage} Alice samples random group elements $g_1,\cdots,g_n$, and constructs the states $|\psi_{g_1}\rangle,\cdots,|\psi_{g_n}\rangle$, which she sends to Bob. 
    \item Bob chooses a random subset $S\subseteq[n]$ of size $n/2$, and sends $S$ to Alice over the classical authenticated channel.
    \item Alice then sends $(g_i)_{i\in S}$ to Bob over the classical authenticated channel.
    \item\label{step:bobcheck} For each $i\in S$, Bob computes $|\psi_{g_i}\rangle$ using $g_i$ and the group action. He then compares this to the $|\psi_{g_i}\rangle$ he received from Alice using the SWAP test. If any of the SWAP tests fail, Bob immediately aborts and rejects.
    \item\label{step:bobsfinalmessage} If Bob does not reject, he does the following for each $i\in [n]\setminus S$. He samples a random $h_i$ and constructs the state $(|\psi_{h_i}\rangle)_{i\in S}$. He also samples a random bit $b_i$. If $b_i=0$, Bob constructs $|\psi_{u_i}\rangle$ where $u_i=g_i+h_i$ by acting on Alice's $|\psi_{g_i}\rangle$ using $h_i$. If $b_i=1$, Bob constructs $|\psi_{u_i}\rangle$ for a random group element $u_i$.

    For each $i\in [n]\setminus S$, Bob sends $|\psi_{h_i}\rangle$ and $|\psi_{u_i}\rangle$ to Alice. Bob outputs $k=(b_i)_{i\in[n]\setminus S}$ as his key.
    \item For each $i\in [n]\setminus S$, Alice acts on $|\psi_{h_i}\rangle$ with $g_i$ to compute $|\psi_{u_i'}\rangle$ where $u_i'=g_i+h_i$. Then Alice performs the SWAP test between $|\psi_{u_i'}\rangle$ and $|\psi_{u_i}\rangle$. If the SWAP test passes, she sets $b_i'=0$. If the SWAP test fails, she sets $b_i'=1$.

    Alice outputs $k'=(b_i')_{i\in[n]\setminus S}$ as her key.
\end{enumerate}   
\end{construction}

At the end of the protocol Bob has a random key $k$, and Alice has a key $k'$. If the adversary did not tamper with the un-authenticated communication, then $k'$ is 0 everywhere that $k$ is 0, and $k'$ is a random bit everywhere that $k$ is 1. Thus $k$ and $k'$ agree in roughly 3/4 of positions. Through standard information-reconciliation techniques, Alice and Bob can securely establish a common shared key.

We now want to argue that Bob's key $k$ is unknown to an attacker that may eavesdrop on the classical authenticated channel and arbitrarily tamper with the un-authenticated channel. First, notice that we do not care about any tampering in Bob's final message in Step~\ref{step:bobsfinalmessage}: this will perturb Alice's key $k'$ but will not affect the adversary's knowledge of $k$. If the adversary perturbs $k'$, this can be remedied through information reconciliation.

If an adversary does not tamper Alice's message in Step~\ref{step:alicefirstmessage}, then for each $i\in[n]\setminus S$, the adversary sees $|\psi_{g_i}\rangle$, $|\psi_{h_i}\rangle$, and $|\psi_{u_i}\rangle$ where $u_i=g_i+h_i$ or random, with $b_i$ indiciating which is the case. Thus, by the DDH assumption, $b_i$ is hidden from the adversary. Note that in the case $|\psi_{g_i}\rangle$ is actually a quantum state, technically the adversary will not even have $|\psi_{g_i}\rangle$ anymore since this was sent to Bob. But DDH still implies security. In the case of quantum information, we can use quantum state group actions with information-theoretic security for DDH (note that orthogonality is not needed for this protocol). In the case of classical communication, we can use any classical group action with computational security for DDH.

If the adversary does tamper with Alice's first message in Step~\ref{step:alicefirstmessage}, then she may have inserted her own $|\psi_{g_i'}\rangle$ into a few of the positions, where the adversary knows $g_i'$. In this case, assuming $i\notin S$ and Bob does not reject during his check in Step~\ref{step:bobcheck}, Alice will actually be able to decode the bit $b_i$ exactly as Alice would. However, any such perturbation to Alice's first message would cause a non-trivial probability that Alice is caught in Bob's check. A standard analysis shows that, conditioned on Bob's check passing, Alice can only learn a small fraction of the $b_i$. Through standard privacy amplification techniques, Alice and Bob can extract uniform keys hidden from the adversary.

%% file: appendix.tex
\appendix